\documentclass[11pt]{article}
\usepackage[utf8]{inputenc}
\usepackage{tikz} 
\usepackage{graphicx}
\usepackage{array}
 \usepackage{setspace}
\usepackage{latexsym}
\usepackage[colorlinks, citecolor = blue ]{hyperref}
\usepackage{algpseudocode}
\usepackage{amsmath}
\usepackage{mathrsfs}
\usepackage{authblk}
\usepackage[linesnumbered,boxed]{algorithm2e}
\usepackage{float}
\usepackage{amssymb}
\usepackage{mdframed}
\newtheorem{theorem}{Theorem}
\newtheorem{asu}{Assumption}
\newtheorem{remark}{Remark}
 
\newtheorem{lem}{Lemma}
\usepackage{amsfonts}
\usepackage{hyperref}
\usepackage[a4paper, total={6.5in, 8.9in}]{geometry}
\usepackage{bbm}
\usepackage{natbib}
\usepackage{mathtools}
\usepackage{graphicx}
\usepackage{placeins}

\numberwithin{equation}{section}
\title{ Kernel Three Pass Regression Filter}
\author{
    Rajveer Jat\thanks{rjat001@ucr.edu. Department of Economics, University of California, Riverside, CA 92507.}
    \quad
    and
    \quad
    Daanish Padha\thanks{Daanish.Padha@glasgow.ac.uk. Adam Smith Business School, University of Glasgow, University Avenue, Glasgow, G12 8QQ}
}

\date{\today}
\begin{document}
\maketitle

\begin{abstract}
We forecast a single time series using a high-dimensional set of predictors. When predictors share common underlying dynamics, a latent factor model estimated by the Principal Component method effectively characterizes their co-movements. These latent factors succinctly summarize the data and aid in prediction, mitigating the curse of dimensionality. However, two significant drawbacks arise: (1) not all factors may be relevant, and utilizing all of them in constructing forecasts leads to inefficiency, and (2) typical models assume a linear dependence of the target on the set of predictors, which limits accuracy. We address these issues through a novel method: Kernel Three-Pass Regression Filter. This method extends a supervised forecasting technique, the Three-Pass Regression Filter, to exclude irrelevant information and operate within an enhanced framework capable of handling nonlinear dependencies. Our method is computationally efficient and demonstrates strong empirical performance, particularly over longer forecast horizons. \footnote{We are grateful to the Editor, the three anonymous referees, and the participants at the 2024 California Econometrics Conference, the 2025 World Congress of the Econometric Society in Seoul, the 34\textsuperscript{th} Midwest Econometrics Group Conference in Kentucky (USA), the 2024 European Winter Meeting of the Econometric Society in Palma de Mallorca (Spain), and the seminar at the University of California, Riverside, for their valuable comments and suggestions.We also thank the Econometric Society for selecting our paper among the top 10\% of contributions presented at the 2025 World Congress for the USD 2,000 fellowship award.} 
\end{abstract}

\section{Introduction}
In recent years, the surge in high-dimensional datasets across fields like economics has ushered in new opportunities and challenges. A paramount issue is the `curse of dimensionality,' which undermines the effectiveness of traditional finite-dimensional estimation methods. Most modeling techniques applied to high-dimensional data assume the existence of a low-dimensional structure that effectively summarizes the data. One stylized feature of high-dimensional economic datasets is the presence of high and pervasive collinearity among variables, leading researchers to posit a data-generating process that assumes all variables are a function of a few latent factors. This formulation is commonly referred to as the factor model. 
A vast amount of literature focuses on using this latent factor structure for forecasting applications. A typical example is found in diffusion index models (\cite{stock2002diff}), where latent factors are derived from a high-dimensional set of variables using Principal Components (hereafter, PC) method. These factors are subsequently utilized to forecast a target variable. A limitation of this PC-based factor estimation is its unsupervised nature, i.e., no information from the target variable is incorporated.

Since the primary goal is to forecast a target rather than estimate the underlying factor structure, introducing a degree of supervision can be beneficial. This can help filter out irrelevant information from the predictor set, thus enhancing the predictive accuracy. This can be done in different ways: using soft and hard thresholding methods to remove predictors with no predictive content, as in \cite{bai2008}, or assigning varying weights to predictors based on their predictive capabilities for the target (see, for example, \cite{scaledpca}), or estimate the subset of factors that exhibit predictive power for the target rather than the complete set of factors that drive the predictors, as in \cite{kelly}. 

The aforementioned models, though offering a supervised alternative to the unsupervised nature of Principal Components, nevertheless rest on the convenient assumption of linearity. However, as underscored in \cite{ml_jae}, non-linearity often characterizes many predictive relationships in economics, particularly over extended time horizons and within data-rich environments.

Various approaches have been proposed to integrate non-linearity into factor models. For instance, squared principal components (PCs) or principal component squared ($PC^2$)  as seen in \cite{bai2008}, sufficient forecasting by \cite{fanSDR}, the kernel trick to estimate factors (\cite{Varlam}) among others. However, these approaches have limited supervision in the prediction process, if any. For example, \cite{fanSDR} estimates factors through an unsupervised method (PC) and then derives sufficient indices using these PCs. Similarly, \cite{Varlam} essentially applies kernel PCA (an unsupervised method) to estimate the set of factors driving a higher-dimensional space obtained by lifting the set of predictors through the kernel method. In \cite{bai2008}, a very particular form of non-linearity (quadratic) is examined, which is somewhat ad hoc. Although they employ thresholding methods to reduce predictors to a smaller set, their screening method, however, may still encounter challenges in filtering relevant factors within this subset, leading to inefficient forecasts.

Our paper incorporates both non-linearity and supervision by introducing a novel kernel three-pass regression filter. The approach applies the three-pass regression filter of \cite{kelly} to a non-linear transformation of the predictor set. While we adopt the lifting idea used in \cite{Varlam}, we depart from unsupervised techniques such as kernel PCA and instead employ a supervised procedure to estimate factors obtained from non-linear transformations of the predictors that are relevant for the target variable. Our results align with the findings in \cite{ml_jae}; over long horizons, where non-linear predictive relations become more pronounced, methods that capture non-linearity, such as ours, consistently outperform their linear counterparts. However, other non-linear approaches do not perform as well. Although they display the expected pattern of improving relative to linear models at longer horizons, the gains are substantially smaller. We attribute the superior performance of our approach to the combination of non-linearity and supervision. Importantly, the non-linear gains at longer horizons do not come at the cost of short-horizon underperformance; while it may not always be the best at very short horizons, it remains close to the best and in several cases even surpasses competing methods. 



The paper proceeds as follows. Section 2 provides a brief introduction to Kernel methods. Section 3 introduces our estimator and discusses its similarity with the estimator of \cite{kelly}. We also list a set of assumptions that ensure the theoretical properties of our estimators, which are given in the subsequent section 4. We present our empirical results in sections 5 and 6 and conclude in section 7. Mathematical proofs and implementation details are given in the appendix.     

\subsection*{Definitions and notations}
We use $\boldsymbol{y}$ to denote the $T \times 1$ vector of the target variable, i.e.
\[
\boldsymbol{y} = (y_h, y_{h+1}, \dots, y_{T+h})'.
\]
We have $N$ predictors with $T$ observations for each predictor. The cross section of predictors at time $t$ is given by the $N \times 1$ vector $\boldsymbol{x}_t$, while the temporal observations of predictor $i$ are given by the vector $\mathbf{x}_i \in \mathbb{R}^T$. Stacking the predictors yields the $T \times N$ matrix
\[
\boldsymbol{X} = (\boldsymbol{x}_1, \boldsymbol{x}_2, \dots, \boldsymbol{x}_T)' = (\mathbf{x}_1, \mathbf{x}_2, \dots, \mathbf{x}_N),
\]
so that the $t$-th row of $\boldsymbol{X}$ is the cross section $\boldsymbol{x}_t' \in \mathbb{R}^{1 \times N}$, and the $i$-th column contains the temporal observations $\mathbf{x}_i \in \mathbb{R}^{T \times 1}$ of predictor $i$.

\noindent We also have $L < \infty$ proxies, collected in the $T \times L$ matrix
\[
\boldsymbol{Z} = (\boldsymbol{z}_1, \boldsymbol{z}_2, \ldots, \boldsymbol{z}_T)',
\]
where each $\boldsymbol{z}_t \in \mathbb{R}^{L}$ is the vector of all $L$ proxies observed at time $t$.

We use $\varphi$ to denote a transformation $\varphi:\mathbb{R}^N\to\mathbb{R}^M$, which transforms each $\boldsymbol{x}_t$, at time $t$, into an $M\times 1$ vector $\varphi(\boldsymbol{x}_t)$. Applying $\varphi$ to the vector of predictors at each time point and stacking the resulting vectors yields the transformed predictor matrix,
\[
\varphi(\boldsymbol{X}) = \big(\varphi(\boldsymbol{x}_1), \varphi(\boldsymbol{x}_2), \dots, \varphi(\boldsymbol{x}_T)\big)' = \big(\varphi_1(\mathbf{x}), \varphi_2(\mathbf{x}), \dots, \varphi_M(\mathbf{x})\big)\in\mathbb{R}^{T\times M},
\]
where the $t$-th row of $\varphi(\boldsymbol{X})$ is the transformed cross section $\varphi(\boldsymbol{x}_t)'\in\mathbb{R}^{1\times M}$ at time $t$, and the $j$-th column $\varphi_j(\mathbf{x})\in\mathbb{R}^{T\times 1}$ represents the vector of temporal values of the $j$-th transformed feature  across times $\{1, \dots, T\}$.  $\boldsymbol{I}_L$ denotes an identity matrix of dimesnion $L$. We use $\boldsymbol{J}_T$ to denote the demeaning matrix, i.e., $\boldsymbol{J}_T \equiv \boldsymbol{I}_T - \frac{1}{T}\mathbf{\iota}_T\mathbf{\iota}_T'$, and for conformable matrices $\boldsymbol{U}$ and $\boldsymbol{V}$ we define $
\boldsymbol{S}_{UV} \equiv \boldsymbol{U}' \boldsymbol{J}_T \boldsymbol{V}.
$. Stochastic orders are denoted by the usual ${O}_{p}$ and ${o}_{p}$. For a matrix, $\boldsymbol{O}_{p}$ and $\boldsymbol{o}_{p}$ denotes the element wise stochastic order, i.e., a matrix is said to be $\boldsymbol{O}_{p}(1)$ or $\boldsymbol{o}_{p}(1)$ if all it's elements are ${O}_{p}(1)$ or ${o}_{p}(1)$ respectively.

\section{Kernel Method} \label{sec2}

We consider a {transformation} 
\(\varphi: \mathbb{R}^N \to \mathbb{R}^M\) 
that maps the original set of predictors \(\boldsymbol{x}_t\) into a vector \(\varphi(\boldsymbol{x}_t)\) in a (typically higher- or potentially infinite-) dimensional space. These transformed vectors are often called features. Stacking the transformed predictors across all times yields the matrix
\[
\varphi(\boldsymbol{X}) = \big(\varphi(\boldsymbol{x}_1), \varphi(\boldsymbol{x}_2), \dots, \varphi(\boldsymbol{x}_T)\big)' \in \mathbb{R}^{T \times M}.
\]

Methods such as principal components and 3PRF, depend on the predictors only through the \(T \times T\) matrix of dot products \(\boldsymbol{X}\boldsymbol{X}'\). Extending these methods to the transformed set of predictors would require computing the Gram matrix 
\(\varphi(\boldsymbol{X}) \varphi(\boldsymbol{X})'\), 
which can be computationally expensive when \(M\) is large or infeasible if the transformed features lie in an infinite-dimensional space.

To address this, we introduce a kernel function $\mathcal{K}(\boldsymbol{x}_t, \boldsymbol{x}_s)$, which computes the inner product of the transformed predictors indirectly as  
\begin{equation} \label{eq:kernel_innerproduct}
\mathcal{K}(\boldsymbol{x}_t, \boldsymbol{x}_s) =  \varphi(\boldsymbol{x}_t)' \varphi(\boldsymbol{x}_s) .
\end{equation}
This approach, known as the `kernel trick', allows us to compute the inner products in the feature space without explicitly evaluating the transformation $\varphi$.  
Stacking these evaluations across all times yields the kernel (Gram) matrix  
\[
\mathcal{K}(\boldsymbol{X}, \boldsymbol{X}') = \varphi(\boldsymbol{X}) \varphi(\boldsymbol{X})' \in \mathbb{R}^{T \times T},
\]
whose $(t,s)$-th entry corresponds to $\mathcal{K}(\boldsymbol{x}_t, \boldsymbol{x}_s)$ as defined in \eqref{eq:kernel_innerproduct}.  
A function $\mathcal{K}$ is called a valid kernel if, for any collection of inputs $\{\boldsymbol{x}_1, \dots, \boldsymbol{x}_T\}$, the resulting matrix $\mathcal{K}(\boldsymbol{X}, \boldsymbol{X}')$ is symmetric and positive semi-definite.  
By Mercer’s theorem (see Appendix~\ref{apx_mercer_theorem}), every such valid kernel is associated with a Hilbert space $\mathcal{H}$ and a mapping $\varphi: \mathbb{R}^N \to \mathcal{H}$ such that the equality in \eqref{eq:kernel_innerproduct} holds.  
In other words, a valid kernel can always be interpreted as the inner product between transformed observations in some  feature space.\footnote{In the Online Appendix-B.1, we illustrate how different kernel functions correspond to the inner products of transformed inputs.}
\begin{figure}[!h]
    \centering
   \includegraphics{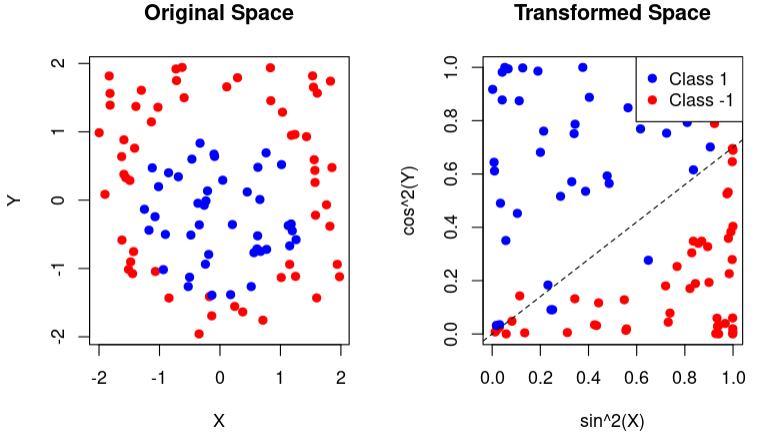}
    \caption{Non-Linear Transformation Making Classification Easy } 
    \label{fig_sin_cos_space}
\end{figure}

Working in a transformed space provides a substantial advantage: many nonlinear relationships among the original predictors can become linear in the appropriately transformed set of predictors. As a stylized example, consider the illustration in Figure~\ref{fig_sin_cos_space}. We generate two variables $X$ and $Y$ from the uniform distribution $U[-2,2]$ and define a binary variable $z$ as
\[
z = 
\begin{cases}
1 & \text{if } X^2 + Y^2 \leq 2, \\
-1 & \text{otherwise.}
\end{cases}
\]
In the original predictor space (Figure~\ref{fig_sin_cos_space}, left), the two classes of $z$ cannot be separated by a linear boundary. However, after transforming the inputs to $\varphi_1(X) = \sin^2(X)$ and $\varphi_2(Y) = \cos^2(Y)$, the classes become linearly separable in the transformed feature space (Figure~\ref{fig_sin_cos_space}, right), where blue and red points represent the two classes. Although this example involves only a two-dimensional input $\boldsymbol{W} = (X, Y)$ transformed into a two-dimensional feature space, in practice $\varphi(\boldsymbol{X})$ can be high- or even infinite-dimensional. Such transformations typically introduce a large set of nonlinear basis functions of the original predictors, making it more likely that an underlying nonlinear relationship in the data becomes linearly representable in the transformed space.

\section{The Estimator}
We delineate the three regression passes that we use to construct our forecast. The first two passes, as explained below, are not feasible in practice, whilst the eventual closed-form solutions are. Nonetheless, these steps offer valuable insights into the underlying process of our estimator and elucidate its similarity to the well-known linear three-pass filter proposed by \cite{kelly}.

 Below, we list the data generation process for the transformed predictor set ($\varphi(\boldsymbol{X})$), the target  ($\boldsymbol{y}$), and the proxies employed for supervision ($\boldsymbol{Z}$). Given the data structure, it is easy to explain why this supervised methodology is effective in estimating the target relevant factors. 

\begin{asu} \label{asu1}
    Data generating Process. \begin{align*}
   \varphi(\boldsymbol{x}_t) & =\boldsymbol{\Phi} \boldsymbol{F}_t+\boldsymbol{\varepsilon}_t &  
    y_{t+h} & =\beta_0+\boldsymbol{\beta}^{\prime} \boldsymbol{F}_t  + \eta_{t+h} &
      \boldsymbol{z}_t & ={\boldsymbol{\lambda}_0}+\boldsymbol{\Lambda} \boldsymbol{F}_t   + \boldsymbol{\omega}_t \\
    \varphi(\boldsymbol{X}) & =\boldsymbol{F} \boldsymbol{\Phi}^{\prime}+\boldsymbol{\varepsilon} 
    & \boldsymbol{y} &=\boldsymbol{\iota}_T \beta_0+\boldsymbol{F} {\boldsymbol{\beta}} + \boldsymbol{\eta} & 
    \boldsymbol{Z}&=\boldsymbol{\iota}_T \boldsymbol{\lambda}_0^{\prime}+\boldsymbol{F} {\boldsymbol{\Lambda}}^{\prime} + \boldsymbol{\omega} 
\end{align*}
where $\boldsymbol{F}_t=\left(\boldsymbol{f}_t^{\prime}, \boldsymbol{g}_t^{\prime}\right)^{\prime}, 
\boldsymbol{\Phi}=\left(\boldsymbol{\Phi}_f, \boldsymbol{\Phi}_g\right), \boldsymbol{\Lambda}=\left(\boldsymbol{\Lambda}_{f}, \boldsymbol{\Lambda}_{g}\right)$, and $\boldsymbol{\beta}=\left(\boldsymbol{\beta}_{f}^{\prime}, \mathbf{0}^{\prime}\right)^{\prime}$ with $\left|\boldsymbol{\beta}_f\right|>\mathbf{0}$. $ K_f>0$ is the dimension of vector $\boldsymbol{f}_t, K_g \geq 0$ is the dimension of vector $\boldsymbol{g}_t, L>0$ is the dimension of vector $\boldsymbol{z}_t$, 
and $K= K_f+ K_g$.
\end{asu}

\noindent The mapping $\varphi$ transforms the original $N$-dimensional predictors $\boldsymbol{x}_t$ into an $M$-dimensional set of features, where $M$ depends on the input dimension $N$. In general, for any (fixed) valid kernel, $M$ increases with $N$, meaning that adding more predictors weakly increases the dimension of the (implied) transformed feature space for a given Kernel. For example, under a second-degree polynomial kernel, the transformed feature vector includes all original linear terms, all squared terms, and all pairwise products of the predictors, so that the total number of features is $M = N + N(N+1)/2$, which grows with $N$.\footnote{See the Online Appendix for more details.} 


Assumption \ref{asu1} endows the transformed set of predictors with a factor structure. 
The coefficient vector $\boldsymbol{\beta}_f$ corresponds to the relevant factors $\boldsymbol{f}_t$ 
in the main predictive equation for $y_{t+h}$; being nonzero, it indicates that these factors directly 
influence the outcome, whereas the coefficients on the irrelevant factors $\boldsymbol{g}_t$ are zero. 
The dimensions $K_f$, $K_g$, and $L$ are all fixed and remain small relative to $N$. 

The factor loading matrix for the transformed predictors $\boldsymbol{\Phi}$ has dimension $M \times K$, mapping the $K$ latent 
factors to the $M$ transformed features. Individual factor loadings $\boldsymbol{\phi}_k \in \mathbb{R}^M$ 
correspond to the weights of factor $k$ across the $M$ transformed features, with 
$\boldsymbol{\Phi} = [\boldsymbol{\phi}_1, \dots, \boldsymbol{\phi}_K]$. The submatrix 
$\boldsymbol{\Phi}_f$ is $M \times K_f$ and contains the loadings on the relevant factors, while 
$\boldsymbol{\Phi}_g$ is $M \times K_g$ and contains the loadings on the irrelevant factors. Similarly, the loading matrix for the proxies, $\boldsymbol{\Lambda}$, has dimension $L \times K$, mapping 
the $K$ latent factors to the $L$ proxy variables. Its submatrices $\boldsymbol{\Lambda}_f$ and 
$\boldsymbol{\Lambda}_g$ have dimensions $L \times K_f$ and $L \times K_g$, respectively, and are 
defined analogously, capturing the contribution of the relevant and irrelevant factors to each proxy.

The proxy variables $\boldsymbol{z}_t$ are generated by the underlying factors together with 
idiosyncratic proxy noise. Overall, the framework parallels \cite{kelly}, with the key distinction 
that the factor structure is applied to transformed (and potentially nonlinear) features rather than 
the raw predictor space.

\noindent Our estimator is characterized by three infeasible three-passes, which are summarized in Table \ref{tab1}. 

\begin{table}[H] 
\centering
   \begin{tabular}{ll} 
\hline Stage-1 \\
\hline Pass & Description \\
\hline 1. & Run time series regression of $\varphi_j(\mathbf{x})$ on $\boldsymbol{Z}$ for $j=1, \ldots, M$, \\
& $\varphi_j(\boldsymbol{x}_t)= \tilde{\phi}_{0, j}+\boldsymbol{z}_{t}^{\prime} \tilde{\boldsymbol{\phi}}_{j}+ \hat{v}_{1 j t}$, retain slope estimate $\tilde{\boldsymbol{\phi}}_{j}$. \\
2. & Run cross section regression of $\varphi(\boldsymbol{x}_{t})$ on $\tilde{\boldsymbol{\phi}}$ for $t=1, \ldots, T$, \\
& $\varphi_j(\boldsymbol{x}_t)= \tilde{\boldsymbol{\phi}}_{j}^{\prime} \hat{\boldsymbol{F}}_{t}+ \hat{v}_{2 j t}$, retain slope estimate $\hat{\boldsymbol{F}}_{t}$. \\
3. & Run time series regression of $y_{t+h}$ on predictive factors $\hat{\boldsymbol{F}}_{t}$, \\
& $\hat{y}_{t+h}=\hat{\beta_{0}}+\hat{\boldsymbol{F}}^{\prime} \hat{\boldsymbol{\beta}}$, delivers the forecast. \\
\hline
\end{tabular} 
\caption{Kernel 3PRF}
\label{tab1}
\end{table}

\begin{figure}[h]
    \centering
    
    \begin{tikzpicture}[scale=0.8,x=3cm, y=2cm, >=stealth]

    \foreach \m/\l [count=\y] in {1,2,...,3}
        \node [every neuron/.try, neuron \m/.try, fill=red!30, minimum size=1cm] (additional-\m) at (2,-3.5-\y) {$z^{(\l)}$};

    \foreach \m/\l [count=\y] in {1,2,3,4}
        \node [every neuron/.try, neuron \m/.try, fill=blue!30] (input-\m) at (0,0.5-\y) {$X_{\l}$};

    \foreach \m [count=\y] in {1,2,3,4}
        \node [every neuron/.try, neuron \m/.try, fill=green!30] (transformed-\m) at (2,0.5-\y) {$\varphi(X_{\y})$};

    \foreach \m [count=\y] in {1,2,...,3}
        \node [every neuron/.try, neuron \m/.try, fill=orange!30] (factors-\m) at (4,-0.5-\y*1.25) {$f^{(\m)}$};

    \node [every neuron/.try, neuron 1/.try, fill=purple!30] (output) at (6,-2.75) {$y$};

    \foreach \i in {1,...,3}
        \foreach \j in {1,...,3}
            \draw [->] (additional-\i) -- (factors-\j);

    \foreach \i in {1,...,4}
        \draw [->] (input-\i) -- (transformed-\i);

    \foreach \i in {1,...,4}
        \foreach \j in {1,...,3}
            \draw [->] (transformed-\i) -- (factors-\j);

    \foreach \i in {1,...,3}
        \draw [->] (factors-\i) -- (output);

    \end{tikzpicture}
    \caption{Implementation of the Three Pass regression filter for the case T=4 and L=3 relevant factors. The variables $z^{(1)} \dots z^{(3)}$ and $f^{(1)} \dots f^{(3)}$ are the vectors representing the time series of the respective variables. $X_s$, (resp $\varphi(X_s))$ represents the cross section of $\boldsymbol{X}$ (resp $\varphi(\boldsymbol{X})$) in period $s$.   }
\end{figure}

These three passes rely on the fact that the correlation between the transformed $\varphi(\boldsymbol{X})$ and the proxies is only due to target relevant factors. Therefore, pass 1 of the regression asymptotically yields a rotation of the relevant-factor loadings of the $j^{th}$ predictor. Cross-sectional covariance between these loadings and the predictors, across $t$, is solely affected by the target relevant factor(s). Hence, pass 2 of this process traces the factor(s) out as a slope parameter. The last pass involves regressing the target variable on the estimated factor(s). Although these three passes offer valuable insights into the mechanics of our process, they are infeasible in practice due to the unavailability of the transformed inputs $\varphi(\boldsymbol{X})$. This is where the kernel trick proves to be useful. To see this, we note that factor(s), their predictive coefficients, and the forecast can be expressed in closed form as below,

\noindent The estimated factor(s) :
\begin{align*}
\hat{\boldsymbol{F}}^{\prime} &=\boldsymbol{S}_{Z Z}\left(\boldsymbol{S}_{\varphi(X) Z}^{\prime} \boldsymbol{S}_{\varphi(X) Z}\right)^{-1} \boldsymbol{S}_{\varphi(X) Z}^{\prime} \varphi(\boldsymbol{X})^{\prime} \\
& = \boldsymbol{Z}^{\prime} \boldsymbol{J}_{T} \boldsymbol{Z}\left(\boldsymbol{Z}^{\prime} \boldsymbol{J}_{T} \varphi(\boldsymbol{X}) \varphi(\boldsymbol{X})^{\prime} \boldsymbol{J}_{T} \boldsymbol{Z}\right)^{-1} \boldsymbol{Z}^{\prime} \boldsymbol{J}_{T} \varphi(\boldsymbol{X}) \varphi(\boldsymbol{X})^{\prime}\\
& =  \boldsymbol{Z}^{\prime} \boldsymbol{J}_{T} \boldsymbol{Z}\left(\boldsymbol{Z}^{\prime} \boldsymbol{J}_{T} \mathcal{K}(\boldsymbol{X},\boldsymbol{X}^{\prime}) \boldsymbol{J}_{T} \boldsymbol{Z}\right)^{-1} \boldsymbol{Z}^{\prime} \boldsymbol{J}_{T} \mathcal{K}(\boldsymbol{X},\boldsymbol{X}^{\prime})
\end{align*}

\noindent The estimated coefficient(s) of the factor(s) :
\begin{align*}
\hat{\boldsymbol{\beta}}&=\boldsymbol{S}_{Z Z} \boldsymbol{S}_{\varphi(\boldsymbol{X}) Z} \boldsymbol{S}_{\varphi(\boldsymbol{X}) Z}\left(\boldsymbol{S}_{\varphi(\boldsymbol{X}) Z}^{\prime} \boldsymbol{S}_{\varphi(\boldsymbol{X}) \varphi(\boldsymbol{X})} \boldsymbol{S}_{\varphi(\boldsymbol{X}) Z}\right)^{-1} \boldsymbol{S}_{\varphi(\boldsymbol{X}) Z}^{\prime} \boldsymbol{S}_{\varphi(\boldsymbol{X}) y}. \\ 
&= \left(\boldsymbol{Z}^{\prime} \boldsymbol{J}_{T} \boldsymbol{Z}\right)^{-1} \boldsymbol{Z}^{\prime} \boldsymbol{J}_{T} \varphi(\boldsymbol{X})  \varphi(\boldsymbol{X})^{\prime} \boldsymbol{J}_{T} \boldsymbol{Z}\left(\boldsymbol{Z}^{\prime} \boldsymbol{J}_{T} \varphi(\boldsymbol{X}) \varphi(\boldsymbol{X})^{\prime} \boldsymbol{J}_{T} \varphi(\boldsymbol{X})  \varphi(\boldsymbol{X})^{\prime} \boldsymbol{J}_{T} \boldsymbol{Z}\right)^{-1} \times \\& \hspace{10mm}\boldsymbol{Z}^{\prime} \boldsymbol{J}_{T} \varphi(\boldsymbol{X})  \varphi(\boldsymbol{X})^{\prime} \boldsymbol{J}_{T} \boldsymbol{y}\\
& = \left(\boldsymbol{Z}^{\prime} \boldsymbol{J}_{T} \boldsymbol{Z}\right)^{-1} \boldsymbol{Z}^{\prime} \boldsymbol{J}_{T} \mathcal{K}(\boldsymbol{X},\boldsymbol{X}^{\prime}) \boldsymbol{J}_{T} \boldsymbol{Z}\left(\boldsymbol{Z}^{\prime} \boldsymbol{J}_{T} \mathcal{K}(\boldsymbol{X},\boldsymbol{X}^{\prime}) \boldsymbol{J}_{T} \mathcal{K}(\boldsymbol{X},\boldsymbol{X}^{\prime}) \boldsymbol{J}_{T} \boldsymbol{Z}\right)^{-1} \boldsymbol{Z}^{\prime} \boldsymbol{J}_{T} \mathcal{K}(\boldsymbol{X},\boldsymbol{X}^{\prime}) \boldsymbol{J}_{T} \boldsymbol{y}
\end{align*}

\noindent Finally, the estimated target :
\begin{align*}
    \hat{\boldsymbol{y}}&= \mathbf{\iota}_{T} \bar{y}+ \boldsymbol{J}_{T}\hat{\boldsymbol{F}} \hat{\boldsymbol{\beta}}\\
    & =\mathbf{\iota}_{T} \bar{y}+\boldsymbol{J}_{T}  \varphi(\boldsymbol{X}) \boldsymbol{S}_{ \varphi(\boldsymbol{X}) Z}\left(\boldsymbol{S}_{ \varphi(\boldsymbol{X}) Z}^{\prime} \boldsymbol{S}_{ \varphi(\boldsymbol{X})  \varphi(\boldsymbol{X})} \boldsymbol{S}_{ \varphi(\boldsymbol{X})  \boldsymbol{Z}}\right)^{-1} \boldsymbol{S}_{ \varphi(\boldsymbol{X}) Z}^{\prime} \boldsymbol{S}_{ \varphi(\boldsymbol{X}) y} \\
    = & \iota \bar{y}+\boldsymbol{J}_{T} \varphi(\boldsymbol{X}) \varphi(\boldsymbol{X})^{\prime} \boldsymbol{J}_{T} \boldsymbol{Z}\left(\boldsymbol{Z}^{\prime} \boldsymbol{J}_{T} \varphi(\boldsymbol{X}) \varphi(\boldsymbol{X})^{\prime} \boldsymbol{J}_{T} \varphi(\boldsymbol{X}) \varphi(\boldsymbol{X})^{\prime} \boldsymbol{J}_{T} \boldsymbol{Z}\right)^{-1} \boldsymbol{Z}^{\prime} \boldsymbol{J}_{T} \varphi(\boldsymbol{X})  \varphi(\boldsymbol{X})^{\prime} \boldsymbol{J}_{T} \boldsymbol{y}\\
    = & \iota \bar{y}+\boldsymbol{J}_{T} \mathcal{K}(\boldsymbol{X},\boldsymbol{X}^{\prime}) \boldsymbol{J}_{T} \boldsymbol{Z}\left(\boldsymbol{Z}^{\prime} \boldsymbol{J}_{T} \mathcal{K}(\boldsymbol{X},\boldsymbol{X}^{\prime}) \boldsymbol{J}_{T} \mathcal{K}(\boldsymbol{X},\boldsymbol{X}^{\prime}) \boldsymbol{J}_{T} \boldsymbol{Z}\right)^{-1} \boldsymbol{Z}^{\prime} \boldsymbol{J}_{T} \mathcal{K}(\boldsymbol{X},\boldsymbol{X}^{\prime}) \boldsymbol{J}_{T} \boldsymbol{y}
\end{align*}

These expressions are obtained by simply replacing $\boldsymbol{X}$ by $\varphi(\boldsymbol{X})$ in the three-pass regression filter of \cite{kelly}.\footnote{The only difference is that the matrix $\boldsymbol{J}_N$, used in \cite{kelly} to demean along the cross-sectional dimension, is absent here. By omitting the intercept in the regression of $\boldsymbol{X}$, we avoid the need for this demeaning step. This is justified because we work with the kernel function rather than explicit transformed features, so the transformation is implicit and passes 1 and 2 are infeasible in practice. Such an approach is standard in kernel-based methods like kernel PCA, where demeaning is performed along the temporal (sample) dimension but not along the transformed predictor dimension.} As evident from the expression of $\hat{\boldsymbol{F}}^{\prime}$, the filtration process applied on the transformed predictor space results in a favorable scenario where the eventual estimate of the factor(s) depends upon the transformed predictors only through their dot products in the transformed space. This holds true for $\hat{\boldsymbol{\beta}}$ and $\hat{\boldsymbol{y}}$ as well.

This inner product can be computed using a suitable kernel function. Alternatively, it can be inferred that employing a positive semidefinite (psd) kernel function to calculate dot products in these derived expressions is akin to executing the three-pass filter process on the transformed set of predictor(s), which, according to Mercer's theorem, are guaranteed to exist.

The key insight is that the variation in the predictors that is relevant for forecasting $y_{t+h}$ may lie on a low-dimensional, nonlinear manifold in the original predictor space. After applying a kernel-induced transformation $\varphi(\cdot)$, this nonlinear structure becomes much closer to linear in the high-dimensional feature space spanned by the transformed features.

By assuming that $\varphi(x_t)$ admits an approximate factor structure, we are effectively stating that the predictive manifold becomes well-approximated by a low-dimensional linear subspace in the transformed feature space. Therefore, while the conditional expectation $E[y_{t+h}\mid x_t]$ may be highly nonlinear in the original $N$-dimensional predictors, the same object becomes linear in a small number of relevant latent factors $f_t$ after the transformation $\varphi(\cdot)$.

This delivers the same idea as \citet{kelly}: forecasting is driven by a few latent factors that summarize the relevant predictive variation. The only difference is that, rather than extracting these factors from the raw predictors, we extract them after a kernel-induced nonlinear transformation, which makes nonlinear predictive structure amenable to linear factor methods.

The Kernel three-pass regression, like the linear 3PRF, relies on the availability of suitable proxies. \cite{kelly} show that such proxies can always be constructed using the target variable $\boldsymbol{y}$. That process is explained in Table-\ref{table_auto_proxy}.

\begin{table}[H]
\centering
\begin{tabular}{p{0.8cm} p{12cm}}
\hline\hline
0. & Initialize $\boldsymbol{r}_0 = \boldsymbol{y}$. For $k = 1, \ldots, L$ (where $L$ is the total number of proxies). \\[0.2cm]

1. & Define the $k^{\text{th}}$ automatic proxy to be $\boldsymbol{r}_{k-1}$.  
Stop if $k = L$; otherwise proceed. \\[0.2cm]

2. & Compute the k3PRF forecast for target $\boldsymbol{y}$ using cross-section $\boldsymbol{X}$ and statistical proxies $1$ through $k$.  
Denote the resulting forecast by $\hat{\boldsymbol{y}}_k$. \\[0.2cm]

3. & Update the residual: $\boldsymbol{r}_k = \boldsymbol{y} - \hat{\boldsymbol{y}}_k$.  
Then increment $k$ and return to Step 1. \\[0.2cm]
\hline\hline
\end{tabular}
\caption{Automatic Proxy-Selection Algorithm}
\label{table_auto_proxy}
\end{table}

\noindent Assumption \ref{asu1} lays out the factor structure of our model. Below, we delineate a set of additional assumptions under which our model delivers consistent forecasts.

\begin{asu} \label{asu2} (Factors, Loadings and Residuals).

Let $R<\infty$. For any $i, s, t$ and some $0<\psi\leq1$,

\begin{enumerate}
    \item $\mathbb{E}\left\|\boldsymbol{F}_{t}\right\|^{4}<R, T^{-1} \sum_{s=1}^{T} \boldsymbol{F}_{s} \underset{T \rightarrow \infty}{\stackrel{p}{\longrightarrow}} \boldsymbol{\mu} \text { and } T^{1/2} \left( \dfrac{\boldsymbol{F}^{\prime} \boldsymbol{J}_{T} \boldsymbol{F}}{T} - \boldsymbol{\Delta}_{F} \right) = \boldsymbol{O}_{p}(1) $. \label{i2.1}

    \item $\mathbb{E}\left\|\phi_{i}\right\|^{4} \leq R$,$ M^{-1} \sum_{j=1}^{M} \phi_{j} \underset{M \rightarrow \infty}{\stackrel{p}{\longrightarrow}} \boldsymbol{0}$, $ M^{1/2}\left( \dfrac{\boldsymbol{\Phi}^{\prime} \boldsymbol{\Phi}}{M}  - \mathcal{P}\right) = \boldsymbol{O}_{p}(1)$. \label{i2.2}

    \item $\mathbb{E}\left(\varepsilon_{i t}\right)=0, \mathbb{E}\left|\varepsilon_{i t}\right|^{8} \leq R$  \label{i2.3}

    \item  $\mathbb{E}\left(\boldsymbol{\omega}_{t}\right)=\mathbf{0}, \mathbb{E}\left\|\boldsymbol{\omega}_{t}\right\|^{4} \leq R, T^{-1 / 2} \sum_{s=1}^{T} \boldsymbol{\omega}_{s}=\boldsymbol{O}_{p}(1)$ and $T^{-1} \boldsymbol{\omega}^{\prime} \boldsymbol{J}_{T} \boldsymbol{\omega} \underset{N \rightarrow \infty}{\stackrel{p}{\longrightarrow}} \boldsymbol{\Delta}_{\omega}$ \label{i2.4}

    \item  $\mathbb{E}_{t}\left(\eta_{t+h}\right)=\mathbb{E}\left(\eta_{t+h} \mid y_{t}, F_{t}, y_{t-1}, F_{t-1}, \ldots\right)=0, \mathbb{E}\left(\eta_{t+h}^{2}\right)=\delta_{\eta}<\infty$, and $\eta_{t+h}$ is independent of $\phi_{i}(m)$ and $\varepsilon_{i, t}$ for any $h>0$. \label{i2.5}
    \end{enumerate} 
\end{asu}

Assumption~\ref{asu2}.\ref{i2.1} imposes regularity on the latent factors by requiring their covariance matrix to converge to a well-behaved, non-stochastic limit, denoted $\boldsymbol{\Delta}_F$, a $K \times K$ matrix. Assumption~\ref{asu2}.\ref{i2.2}, adapted from \cite{kelly}, imposes an analogous condition on the factor loadings, treating $\mathcal{P}$ as a non-stochastic $K \times K$ matrix representing their limiting cross-sectional covariance. Likewise, $\boldsymbol{\Delta}_\omega$ is a non-stochastic $L \times L$ matrix capturing the population covariance of the proxy noise.
  Since we assume a factor structure on the transformed space instead of the original predictor space, the normalization in various terms features $M$ and not $N$, where $M$ is the dimension of our transformed space. Assumptions \ref{asu2}.\ref{i2.3}-\ref{asu2}.\ref{i2.5}, borrowed from \cite{kelly}, impose regularity on various error processes.     

 \begin{asu} \label{asu3} (Dependence).

Let $x(m)$ denote the $m^{\text {th }}$ element of $\boldsymbol{x}$. For $R<\infty$ and any $i, j, t, s, m_{1}, m_{2}$
\begin{enumerate}
    \item $\mathbb{E}\left(\varepsilon_{i t} \varepsilon_{j s}\right)=\sigma_{i j, t s},\left|\sigma_{i j, t s}\right| \leq \bar{\sigma}_{i j}$ and $\left|\sigma_{i j, t s}\right| \leq \tau_{t s}$, and \label{i3.1} \vspace{1mm}\\
    \
        a. $M^{-1} \sum_{i, j=1}^{M} \bar{\sigma}_{i j} \leq R$ \hspace{12mm}
         b. $T^{-1} \sum_{t, s=1}^{T} \tau_{t s} \leq R$   \vspace{1mm}\\
         c. $M^{-1} \sum_{i, s}\left|\sigma_{i i, t s}\right| \leq R$ \hspace{12mm}
         d. $M^{-1} T^{-1} \sum_{i, j, t, s}\left|\sigma_{i j, t s}\right| \leq R$
        
    \item $\mathbb{E}\left|M^{-1 / 2} T^{-1 / 2} \sum_{s=1}^{T} \sum_{i=1}^{M}\left[\varepsilon_{i s} \varepsilon_{i t}-\sigma_{i i, s t}\right]\right|^{4} \leq R$ \label{i3.2}\\

    \item $\mathbb{E}\left|T^{-1 / 2} \sum_{t=1}^{T} F_{t}\left(m_{1}\right) \omega_{t}\left(m_{2}\right)\right|^{2} \leq R$ \label{i3.3}

    \item  $\mathbb{E}\left|T^{-1 / 2} \sum_{t=1}^{T} \omega_{t}\left(m_{1}\right) \varepsilon_{i t}\right|^{2} \leq R$. \label{i3.4}

    \item $\mathbb{E}\left|T^{-1 / 2} \sum_{t=1}^{T} F_{t}\left(m_{1}\right) \varepsilon_{i t}\right|^{2} \leq R$ \label{i3.5}

    \item  $\mathbb{E}\left|M^{-1 / 2} \sum_{i=1}^{M} \phi_{i}\left(m_{1}\right) \varepsilon_{i t}\right|^{2} \leq R$. \label{i3.6}

    \item $\mathbb{E}\left|T^{-1 / 2} \sum_{t=1}^{T} F_{t}\left(m_{1}\right) \eta_{t+h}\right|^{2} \leq R$\label{i3.7}
  
\end{enumerate}
\end{asu}

Assumption \ref{asu3}.\ref{i3.1}-\ref{asu3}.\ref{i3.2} allow various forms of weak cross-sectional and temporal dependence between the idiosyncratic components of the transformed predictors. These assumptions characterize our `Approximate' factor model. The terminology of approximate, as opposed to a strict factor model, alludes to the allowance of these weak correlations, as outlined by \cite{chamberlain}. These assumptions are standard in the literature; see \cite{bai2003}. Assumption \ref{asu3}.\ref{i3.4}-\ref{asu3}.\ref{i3.7} are either borrowed from or are weaker versions of Assumptions in \cite{kelly}. They are reasonable because each of them involves a product of orthogonal series. 

\begin{asu} \label{asu4} (Normalization). 

\begin{enumerate}
    \item $\mathcal{P}=\boldsymbol{I}_K$ \label{i4.1}
    \item $\boldsymbol{\Delta}_{F}$ is diagonal, positive definite, and each diagonal element is unique and bounded. \label{i4.2}
\end{enumerate} 
\end{asu}
Assumption \ref{asu4} characterizes the covariance matrices defined in Assumption \ref{asu2}. This is a normalization assumption and is common in factor model literature, see \cite{kelly} Assumption 5. It pertains to the non-identifiability of the true factor(s). It is well known that only the vector space spanned by the factor(s) can be consistently estimated but not the factor themselves. Imposing some normalization condition for the uniqueness of solution(s) is common in literature.

\begin{asu} \label{asu5} (Relevant Proxies).

\begin{enumerate}
    \item $\boldsymbol{\Lambda}=\left[\begin{array}{ll}\boldsymbol{\Lambda}_{f} & \mathbf{0}\end{array}\right]$ \label{i5.1}
    \item $\boldsymbol{\Lambda}_{f}$ is non-singular. \label{i5.2}
\end{enumerate}
\end{asu}
Assumption \ref{asu5} outlines the role and structure of the proxies in our framework. 
The proxy loading matrix is of the form $\boldsymbol{\Lambda} = [\boldsymbol{\Lambda}_f \;\; \mathbf{0}]$, 
so that the proxies load only on the relevant factors and have zero loading on the irrelevant factors 
($\boldsymbol{\Lambda}_g = \mathbf{0}$). Non-singularity of $\boldsymbol{\Lambda}_f$ ensures that the 
proxies are linearly independent and none are redundant. Implicitly, this also requires that the number 
of proxies $L$ matches the number of relevant factors $K_f$, so that the relevant factor space is 
fully spanned by the proxies. In other words, the common component of the proxies captures all 
variation in the relevant factors, while none of the proxy variation is due to irrelevant factors.

\section{Theoretical Results} 
We show that our estimated forecast converges to the infeasible best in probability. To show the same, we prove some intermediate results. All the proofs are in the appendix. \\[2mm]
Define $\delta_{M T} \equiv min\{ \sqrt{M}, \sqrt{T}\}$.  
\noindent Define $\boldsymbol{H}_{f} \equiv \hat{\boldsymbol{F}}_{A} \hat{\boldsymbol{F}}_{B}^{-1} \mathbf{\Lambda}_f\boldsymbol{\Delta}_{f}$ where,
$\hat{\boldsymbol{F}}_{A} = T^{-1} \boldsymbol{Z}^{\prime} \boldsymbol{J}_{T} \boldsymbol{Z}$ and $\hat{\boldsymbol{F}}_{B} =M^{-1} T^{-2} \boldsymbol{Z}^{\prime} \boldsymbol{J}_{T} \varphi(\boldsymbol{X}) \varphi(\boldsymbol{X}^{\prime}) \boldsymbol{J}_{T} \boldsymbol{Z}$, and $\boldsymbol{\Delta}_{f}$ denotes the submatrix of $\boldsymbol{\Delta}_F$ formed by its first $K_f$ rows and first $K_f$ columns.
\begin{theorem} \label{t1}
 If Assumption \ref{asu1}-\ref{asu5} hold, we have 
    \[
\hat{\boldsymbol{F}}_{t}- \boldsymbol{H}_f \boldsymbol{f}_{t} = \boldsymbol{O}_p(\delta_{M T}^{-1} )
\]
\end{theorem}
This theorem establishes the estimated factor(s) convergence to the true relevant factors up to a rotation. It is well known in the literature on factor models\footnote{This feature of inherent unidentifiability has been emphasized in \cite{bai2003}, \cite{kelly} among other papers. The normalization imposed in assumption \ref{asu5} is done to handle this issue.}, that true underlying factor(s) are not identifiable; we instead estimate a rotated version of the true factors, which preserves their span. 

\vspace{5mm}
\noindent Define $\boldsymbol{G}_{\beta} \equiv \hat{\boldsymbol{\beta}}_{1} ^{-1} {\hat{\boldsymbol{\beta}}_{2}} \left[\boldsymbol{\Lambda}_f \boldsymbol{\Delta}_{f}^3  \boldsymbol{\Lambda}_f^{\prime}\right]^{-1} \boldsymbol{\Lambda}_f \boldsymbol{\Delta}_{f}^2 $, where\\
$\hat{\boldsymbol{\beta}}_{1}=\hat{\boldsymbol{F}}_{A}$ and $\hat{\boldsymbol{\beta}}_{2}=\hat{\boldsymbol{F}}_{B}$, and $\boldsymbol{\Delta}_{f}$ is defined in the paragraph preceding Theorem \ref{t1}. 
\begin{theorem} \label{t2}
 If Assumption \ref{asu1}-\ref{asu5} hold, we have 
   $$\hat{\boldsymbol{\beta}}-\boldsymbol{G}_{\beta}\boldsymbol{\beta}_f =  \boldsymbol{O}_p(\delta_{M T}^{-1} ).$$ 

${\boldsymbol{H}_f}'\boldsymbol{G}_{\beta} = \boldsymbol{I}_{K_f}$
\end{theorem}

This theorem establishes the convergence of the predictive coefficients to a rotation of the true coefficients. Just like in the case of factor(s), true coefficients are not identifiable and we instead estimate their rotation. The rates established in \textbf{Theorems} \ref{t1} and \ref{t2} differ from the rates established in \cite{kelly} and the reason is that our definition of rotation matrices $\boldsymbol{H}_f$ and $\boldsymbol{G}_{\beta}$ are different from \cite{kelly}. (See Remark \ref{r1}).
\begin{remark} \label{r1}
    As highlighted in \cite{bai2006} and also emphasized in \cite{kelly}, the presence of matrices $\boldsymbol{H}_f$ and $\boldsymbol{G}_{\beta}$ in \textbf{Theorem} \ref{t1} and \ref{t2} highlight the fact we are essentially estimating a vector space. These theorems ``pertain to the difference between $\left[\hat{\boldsymbol{F}}_{t} / \hat{\boldsymbol{\beta}}\right]$ and the space spanned by $\left[\boldsymbol{F}_{t} / \boldsymbol{\beta}\right]$". The product $\boldsymbol{H}_f'\boldsymbol{G}_{\beta}$ equals an identity matrix, cancelling the rotations in the estimated coefficients and the factors; thereby consistently estimating direction spanned by $\boldsymbol{\beta}'\boldsymbol{F}_t$. However, this characteristic is absent in Theorems 5 and 6 of \cite{kelly}. The matrices $\boldsymbol{H}$ and $\boldsymbol{G}_{\beta}$ as defined in their paper do not necessarily yield a product that equals an identity matrix. 
\end{remark}

\begin{theorem} \label{t3}
 If Assumption \ref{asu1}-\ref{asu5} hold, we have 
 $$ \hat{y}_{t+h}- \mathbb{E}_{t} y_{t+h} = O_p(\delta_{M T}^{-1}), $$
\end{theorem}
where $\mathbb{E}_{t} y_{t+h} = \beta_0+\boldsymbol{\beta}^{\prime} \boldsymbol{F}_t = \beta_0+\boldsymbol{\beta}_f^{\prime} \boldsymbol{f}_t $.

 Combining \textbf{Theorem} \ref{t1} and \ref{t2}, the convergence of $\hat{y}_{t+h}$ follows directly. Our proof, unlike \cite{kelly} uses the convergence results for the estimated factor(s) and coefficients to obtain this result.  
\begin{remark} \label{rr}
 The rates established in \textbf{Theorem} \ref{t1}, \ref{t2} and \ref{t3} are different from the result in \cite{kelly} where the corresponding rates are $O_p({T}^{-1/2})$, $O_p({T}^{-1/2})$ and $O_p({N}^{-1/2})$\footnote{For our case, it should have been $O_p({M}^{-1/2})$ as per their theorem since we apply 3PRF to the transformed M-dimesnional space.} respectively (see \textbf{Theorems} 4, 5 and 6 in their paper). For \textbf{Theorem} \ref{t1} and \ref{t2}, the difference is explained by a different definition of the rotation matrices in our paper (see \textbf{Remark} \ref{r1}). For establishing the convergence of $\hat{y}_{t+h}$, their proof follows two steps. First they show that $\hat{y}_{t+h}-\tilde{y}_{t+h} = O_p({T}^{-1/2})$, where $\tilde{y}_{t+h}$ is defined in their appendix. Then then they argue that $\sqrt{T} \tilde{y}_{t+h} \underset{T, N \rightarrow \infty}{\longrightarrow} \mathbb{E}_{t} y_{t+h}$. Since $\tilde{y}_{t+h}$\footnote{The definition of $\tilde{y}_{t+h}$ and fact that it is $O_p(1)$ can be seen from the proof of \textbf{Theorem} 4 of \cite{kelly}} is $O_p(1)$,  $\sqrt{T} \tilde{y}_{t+h}$ would diverge to infinity and their statement would be false. We presume that they erroneously wrote this and instead wanted to imply that $\sqrt{T} \left( \tilde{y}_{t+h} - \mathbb{E}_{t} y_{t+h} \right) \underset{T, N \rightarrow \infty}{\longrightarrow} 0$. However this statement is false because $\tilde{y}_{t+h} - \mathbb{E}_{t} y_{t+h} $ has random elements which converge to 0 at a rate which is $O_p(M^{-1/2}) + O_p(T^{-1/2}) = O_p(\delta_{M T}^{-1}).$ 
 \end{remark}

\section{Empirical Applications}
We apply our proposed method to real-world applications, focusing on forecasting time series variables across various economic domains such as national income, finance, labor, housing, prices, etc. To assess the performance of our approach, we conduct comparative analyses against competitive methods, employing the out-of-sample $R^2$ performance metric as a benchmark. The out-of-sample $R^2$ metric is computed as:
\[
R^2 = 1 - \frac{\sum_{i \in \text{test-data}} (y_i - \hat{y}_i)^2}{\sum_{i \in \text{test-data}} (y_i - \bar{y}_{\text{train}})^2}
\]
It computes the out-of-sample proximity of our forecast $\hat{y}$ with the target ($y$) relative to a historical mean $(\bar{y})$; a positive value indicates that the forecast is better than the historical mean. Detailed explanations of performance metrics computation are provided in the Online Appendix-B.2.

\subsection{Overview}
This section provides an introduction to our competing methods and then a high-level summary of the empirical applications. 

\subsubsection{Competing Methods}

We compare our method against eight different forecasting approaches. The first is the principal components (PC) regression proposed by \cite{stock2002diff}, denoted as \textit{PC} in our performance tables. We also consider \textit{PC-Squared} (\textit{PC-Sq}) and \textit{Squared-PC} (\textit{Sq-PC}) from \cite{bai2008}, kernel PCA (\textit{kPCA}) [\cite{kpca}], our linear counterpart \textit{3PRF} (\cite{kelly}), and the autoregressive (AR) model with optimal lag $p$.\footnote{We evaluate several choices of $p$ and report the best-performing one. This represents the ideal performance for AR models in a setting where the true lag is known; in practice, $p$ cannot be chosen based on maximum out-of-sample $R^2$.}

The Diffusion Index (DI) serves as a benchmark forecasting method that combines AR and PCA. We select the number of principal components using the eigenvalue ratio test of \cite{ahn2013eigenvalue} and run several DIs for different AR($p$), retaining the one with the $p$ that achieves the highest out-of-sample $R^2$. In this sense, we assess the upper bound of DI performance, since in practice the optimal $p$ cannot be chosen based on out-of-sample results.

Finally, we consider the Autoencoder (AE), a modern neural network method that learns a latent-space or manifold representation (\cite{AE_hinton}). The crucial hyperparameter, the number of latent dimensions, is chosen via cross-validation, consistent with the hyperparameter tuning applied to the other methods. AE is conceptually the closest neural-network-based competitor to our approach, as it also learns a latent representation of the data. Some of these methods require hyperparameter tuning to achieve optimal results; we perform such tuning as described in subsection \ref{subsec_tuning}.

As discussed in Section \ref{sec2}, different kernel functions correspond to different $\varphi(\cdot)$. We use the Gaussian\footnote{The Gaussian kernel is mathematically equivalent to the commonly used RBF kernel in some papers.} kernel throughout our applications.

\[
K_\sigma(\mathbf{x}, \mathbf{x'}) = \exp \left( -\frac{\|\mathbf{x} - \mathbf{x'}\|^2}{2\sigma^2} \right)
\]

where \(\sigma\) is a hyperparameter determined via cross-validation. The use of this kernel is justified by its strong performance in macroeconomic forecasting, as documented by \cite{rbf_ref1}, \cite{k_ridge}, and \cite{Varlam}. 

\subsubsection{Summary of Empirical Applications}

We begin by discussing the data sources, data transformations for stationary series, and the hyperparameter tuning procedure. The first empirical application illustrates the working of our method using theory-guided proxies. We then provide visual evidence showing how our method captures the nonlinearities present in the target variable compared with its nearest competitor. Next, we present the out-of-sample forecasting performance for several key variables of interest across a broad range of economic categories. Finally, we compare our method against others in a comprehensive forecasting exercise using all available series in the dataset as a target.

\subsection{Data and Hyper-Parameter Tuning}\label{sec_data_tuning}
We utilize the quarterly macroeconomic dataset, FRED-QD. It spans the time period 1959-2023. This dataset encompasses a comprehensive array of more than 250 variables\footnote{There are 176 series for which we can find no missing values from 1959 to 2023, therefore our number of time series used are 176 in this analysis.}, including macroeconomic (such as GDP, Consumption, and Investment), financial, labor market, housing, and industrial and manufacturing variables. We present a tabulation comprising the mnemonic codes and details of the variables in the FRED-QD dataset alongside their counterparts in the Stock-Watson dataset in Online Appendix-B.4 for the series we forecast in this section.

\subsubsection{Data Transformation}
Estimation exercises involving nonstationary time series pose significant challenges. Nonstationarity is ubiquitous in economic and financial data. Nonstationary variables lack a defined population mean, and the sample standard deviation tends to diverge as the number of observations increases, see \cite{Onatski_Wang} for a more detailed discussion.  Generally, researchers address this by manually examining each series to identify necessary transformations before computing principal components.
\cite{hamilton_tranformation} offers an improved method for transforming the predictors to achieve stationarity. We use their method to make our data stationarity. 

Scholars in the literature often employ sample periods devoid of structural breaks. \cite{fan2023latent} notes that ``There exist significant structural breaks for many variables around the year of financial crisis in 2008 which makes our data non-stationary even after performing the suggested transformations''. Therefore, our study focuses on the stationary period spanning from 1965 to 2007\footnote{Another indirect advantage of the choice of this sample period is that it gives us the number of samples less than the number of predictors ($T<N$), hence a truly high-dimensional scenario to test our method in.}. For robustness checks, we examine three other sample periods. First, 1965–2019, which captures the Global Financial Crisis (GFC), allowing us to assess performance during high-volatility periods, as the GFC falls within the test sample under our 70\% training split. Second, 1965–2023, the full sample\footnote{Although our data begins in 1959, we exclude early observations to accommodate AR(p) models with sufficiently large lags.}, encompassing both high- and low-volatility periods, including the GFC, COVID-19, and the Great Moderation. Third, 1984–2007, which corresponds to the Great Moderation and enables the evaluation of performance in a low-volatility environment.  We found that the results are qualitatively similar across all sub-sample periods and convey the same story as in the main text (The results are tabulated in Online Appendix B.9). These findings indicate that our method is robust to both high-volatility (turbulent) and low-volatility (Great Moderation) periods.

In our main analysis, the sample period from 1964:Q4 to 2007:Q1 comprises $T=170$ observations (periods) and $N=176$ variables (predictors). While the data is initially available for around 250 series, those with missing values are removed; this leaves us with a total of 176 series. The model training and hyperparameter tuning are conducted within a rolling window framework, utilizing 70\% of the total observations as the width of the rolling window. We observe qualitatively similar performance across varying window widths (50\%, 60\% of total data). 

\subsubsection{Hyper-parameter Tuning}\label{subsec_tuning}
Our methodology incorporates the kernel as a fundamental element of the estimation process. The kernel function includes a hyperparameter that necessitates optimization. Concurrently, a similar hyperparameter requires tuning in the context of a competitor method, namely kernel PCA. Thus, we employ an identical tuning procedure for both methodologies. We use Gaussian kernel, which relies on a single hyperparameter, denoted as $\sigma$, for our specific applications.  We partition the data into two folds and conduct cross-validation to determine the optimal tuning parameter. Further elaboration on the algorithm employed for this purpose is provided in the Online Appendix-B.5. We need to choose an optimal lag $p$ for the autoregressive (AR) and Diffusion Index (DI) methods, which we have already discussed. For the Autoencoder (AE) method, the key tuning parameters include the number of layers, layer size, and the dimension of the latent space, of which the latent space dimension is the most critical. We conduct experiments by varying the number of layers and layer sizes, and determine the optimal latent space dimension through cross-validation.
Furthermore, a similar cross-validation procedure is applied to determine other hyperparameters in competing methods to ensure a fair comparison.\\[2mm]
Among our competitive methodologies, where factors are computed as PCs, we are required to specify the number of factors.  To address this, we employ the eigenvalue ratio test method proposed by \cite{ahn2013eigenvalue}. This method computes the ratio of each eigenvalue to its predecessor and selects the number of principal components corresponding to the index where this ratio attains its maximum value. We employ a single factor throughout our analyses in both the 3PRF and kernel 3PRF models. This choice is often prudent within the 3PRF setting, as elucidated by \cite{kelly}, who highlight instances where a single factor can effectively represent a multi-factor system. When factors exhibit the same variances\footnote{See appendix section A.7.2 in \cite{kelly}}, a single proxy achieves optimal performance, and even when variances are not identical but closely aligned, one factor estimated through a single auto proxy typically explains a significant portion of the variation\footnote{See simulations in appendix A.7.3 in \cite{kelly}}, rendering residual variation minimal. While we assessed the performance of our estimator with varying numbers of factors, we consistently observed that a single factor predominates; thus, we report results based on this configuration.

\subsection{Forecasting Using Theory Guided Proxies}
The primary objective of this subsection is to establish the viability of theory-guided proxies in forecasting using our method. We also compare with the linear benchmark, i.e., 3PRF of \cite{kelly}. A more extensive performance evaluation will be presented in subsequent subsections,  where we used the auto-proxy method discussed in table-\ref{table_auto_proxy} to construct forecasts using Kernel 3PRF. 

\subsubsection{Forecasting GDP Using Investment and Consumption}
Consumption and investment are well-known drivers of GDP, and they likely capture the underlying latent factors that drive future economic activity. Using these variables as theory-guided proxies allows us to extract the relevant predictive components from the broader cross section of macroeconomic predictors. We construct GDP forecasts using these variables as proxies and report the results in Table-\ref{theory_proxy_GDP}.
\begin{table}[!h]
    \centering
    \begin{tabular}{lccc}
    \hline
    Proxy & 3PRF & k3PRF   \\
    \hline  
     Consumption and Investment  & 0.621& 0.768 \\
     Investment  & 0.627 & 0.748\\
     Consumption  & 0.589 & 0.760\\
\hline 
    \end{tabular}
    
    \caption{One-period Ahead Out-of-Sample $R^2$ for National Income}
    \label{theory_proxy_GDP}
\end{table}
This exercise proves the efficacy of theory-guided proxies. Our method outperforms the nearest competitive method \cite{kelly}.

\subsubsection{Forecasting Inflation using Quantity Theory of Money}
We also reproduce the theory-guided proxy example from \cite{kelly}, focusing on one-period-ahead inflation forecasts. 
The target variable is the change in the price level ($\Delta$Price Level), and the dynamic quantity theory of money relates future inflation to current growth in real output and money supply:

\[
\frac{\Delta(\text{Money supply}) \times \Delta(\text{Velocity of money})}{\Delta (\text{Real Product})} = \Delta (\text{Price level})
\]

Table \ref{Quantity_theory_proxy} reports the results. Using both $\Delta$GDP and $\Delta$Money Supply as proxies 
captures the inflation-relevant variation more effectively than using either alone.

\begin{table}[!h]
    \centering
    \begin{tabular}{lccc}
    \hline
    Proxy &  3PRF & k3PRF   \\
    \hline  
    $\Delta$GDP and $\Delta$Money-Supply & 0.265 & 0.265  \\
    $\Delta$GDP & 0.037 &  0.037\\
    $\Delta$Money Supply & 0.350 & 0.355  \\	
    \hline 
    \end{tabular}
       \caption{One-period Ahead Out-of-Sample $R^2$ for Inflation}
    \label{Quantity_theory_proxy}
\end{table}

The results indicate that the theory-guided proxies successfully capture the key inflation dynamics, achieving performance comparable to the linear benchmark 3PRF. Importantly, this also demonstrates that in settings where the relationship is approximately linear, the kernel 3PRF can adapt and produce forecasts of similar quality, as appropriate tuning of the hyperparameters allows it to effectively capture linear effects. It is important to emphasize again that this analysis focuses on one-step-ahead forecasts, which are not the primary strength of our methodology. The purpose of presenting these results is solely to demonstrate the workings of the procedure through the theory-guided proxies.

\FloatBarrier
\subsection{Comparative Forecasting Plots}\label{forecast_plots}
\begin{figure}[!h]
\centering
\includegraphics[width=150mm]{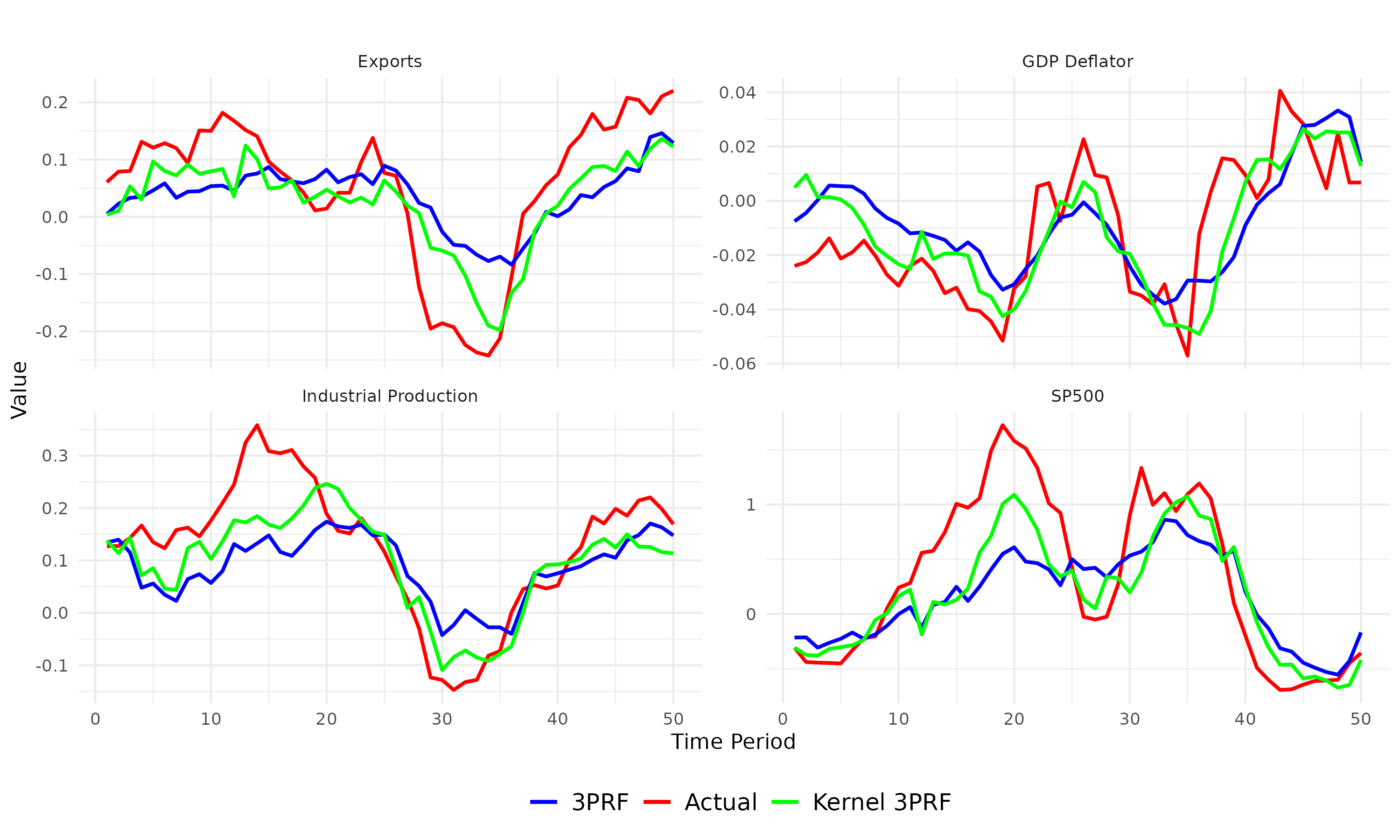}
\caption{Short Horizon (One period ahead) Forecasting: Comparative Performance}
\label{short_run_graph}
\end{figure}

\begin{figure}[!h]
    \centering
    \includegraphics[width=150mm]{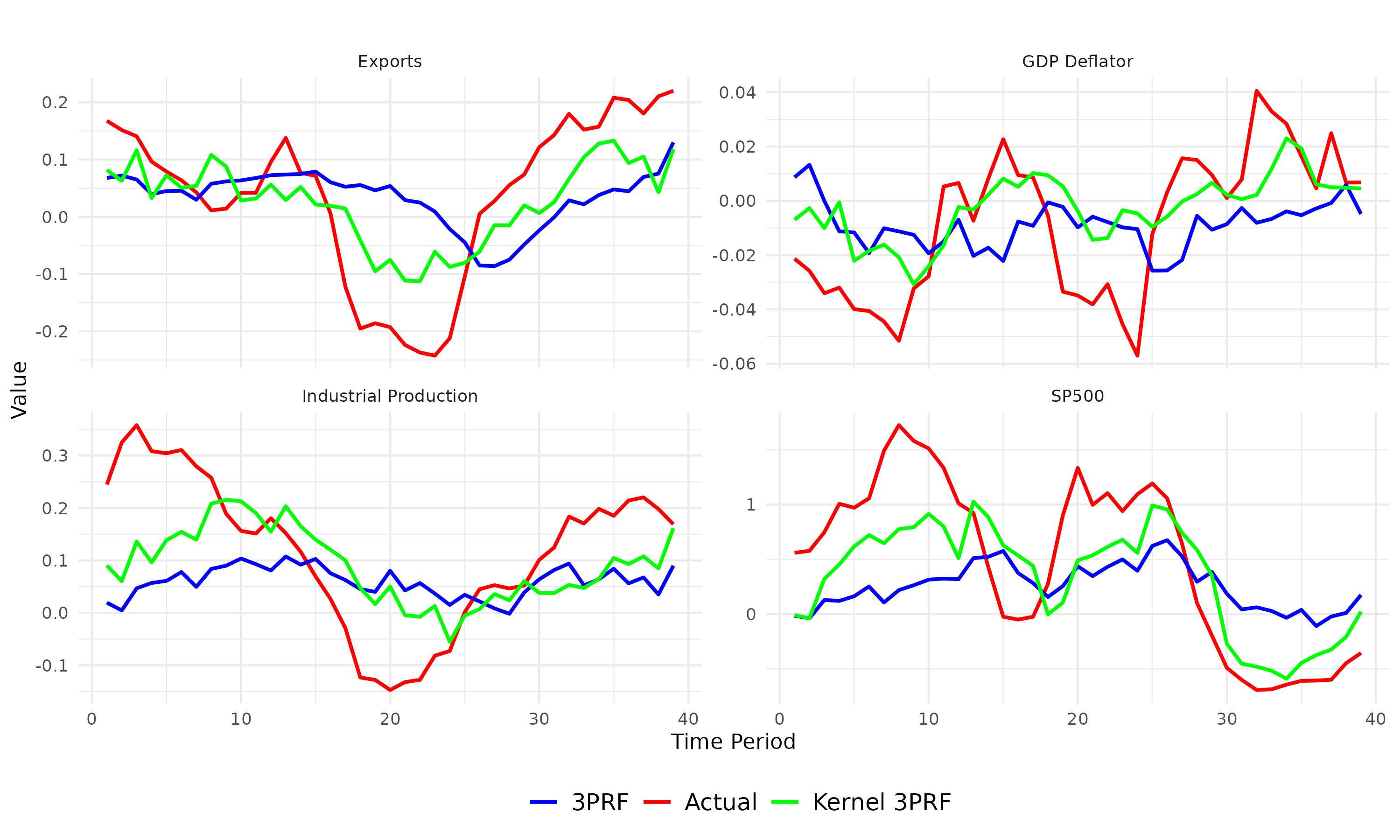}
      \caption{Long Horizon (Twelve periods ahead) Forecasting: Comparative Performance}

    \label{long_run_plot}
\end{figure}

To visually demonstrate the enhanced performance of kernel 3PRF compared to its linear counterpart, we provide comparative performance plots across four distinct types of economic series spanning various domains: macroeconomic series (\textit{Exports}), price series (\textit{GDP Deflator}), manufacturing series (\textit{Industrial Production}), and financial series (\textit{S\&P 500 Index}) in figure-\ref{short_run_graph} and \ref{long_run_plot}. Plots of all other series on different forecast horizons are given in the Online Appendix-B.6.

\FloatBarrier

\subsection{Forecasting Aggregate Macroeconomic Variables}
 An astute economic decision, such as monetary policy formulation, hinges upon well-informed anticipations of future trends in macroeconomic and financial data. Consequently, forecasting macroeconomic variables emerges as a pivotal pursuit for economists. Quoting Federal Reserve of New York's website, \cite{kim_macro_forecasts} notes, ``In formulating the nation’s monetary policy, the Federal Reserve considers a number of factors, including the economic and financial indicators which follow, as well as the anecdotal reports compiled in the Beige Book. Real Gross Domestic Product (GDP); Consumer Price Index (CPI); Nonfarm Payroll Employment Housing Starts; Industrial Production/Capacity Utilization; Retail Sales; Business Sales and Inventories; Advance Durable Goods Shipments, New Orders and Unfilled Orders; Lightweight Vehicle Sales; Yield on 10-year Treasury Bond; S\&P 500 Stock Index; M2''. We, therefore, aim to forecast some of these crucial indicators in this paper. We compare the performance of our model against the competitors. This section forecasts seven macro series: GDP, Consumption, Investment, Exports, Imports, Fixed Investment, and Industrial Production (Final). 

\begin{table}[!h]
    \centering
     \resizebox{130mm}{!}{
   
     \begin{tabular}{llccccccccc}
    \hline \\
\textbf{GDP}& & & & & & &  & & \\     
    \hline
   & Method & h=1 &  h=2 & h=3 & h=4 & h=6 & h=8 & h=10 & h=12 \\ \hline 
& DI & 0.915 & 0.821 & 0.696 & 0.551 & 0.208 & -0.133 & -0.019 & 0.017\\
&AR(p) &0.922 & 0.806 & 0.651 & 0.477 & 0.076 & -0.320 & -0.195 & -0.072\\
&PCA         & 0.717 & 0.650 & 0.575 & 0.492 & 0.311 & 0.130  & -0.001 & -0.075 \\
&Sq-PC  & 0.615 & 0.593 & 0.552 & 0.488 & 0.290 & 0.076  & -0.092 & -0.166 \\
&PC-Sq  & 0.773 & 0.733 & 0.676 & 0.594 & 0.398 & 0.175  & 0.008  & -0.063 \\
&kPCA  & 0.638 & 0.589 & 0.528 & 0.464 & 0.322 & 0.204  & 0.060  & 0.063 \\
& AE & 0.463 & 0.449 & 0.459 & 0.234 & 0.192 & 0.124 & -0.094 & -0.069 \\
&3PRF   & 0.667 & 0.619 & 0.561 & 0.493 & 0.341 & 0.193  & 0.130  & 0.201 \\
&k3PRF & 0.808 & 0.788 & 0.757 & 0.701 & 0.603 & 0.544  & 0.608  & 0.434 \\[1mm]

\textbf{Consumption}& & & & & & &  & & \\     
   \hline
&    Method & h=1 &  h=2 & h=3 & h=4 & h=6 & h=8 & h=10 & h=12 \\ \hline  
& DI & 0.953 & 0.891 & 0.803 & 0.687 & 0.430 & 0.173 & 0.145 & 0.198\\
&AR(p)  &0.953 & 0.879 & 0.776 & 0.643 & 0.352 & 0.068 & 0.080 & 0.152\\
&PCA    & 0.573 & 0.554 & 0.504 & 0.430 & 0.238 & 0.038 & -0.093 & -0.155 \\
&Sq-PC   &0.546 & 0.541 & 0.499 & 0.428 & 0.235 & 0.025 & -0.137 & -0.206 \\
&PC-Sq   &0.611 & 0.637 & 0.628 & 0.596 & 0.412 & 0.161 & -0.041 & -0.128 \\
&kPCA   &0.433 & 0.419 & 0.369 & 0.319 & 0.143 & 0.076 & 0.039  & 0.181 \\
& AE & 0.529 & 0.459 & 0.411 & 0.370 & 0.241 & 0.185 & 0.041 & -0.003 \\
&3PRF         &0.589 & 0.547 & 0.501 & 0.464 & 0.386 & 0.196 & 0.169  & 0.326 \\
&k3PRF   &0.713 & 0.730 & 0.720 & 0.741 & 0.770 & 0.747 & 0.275 & 0.496\\[1mm]

\textbf{Investment}& & & & & & &  & & \\   
\hline
&    Method & h=1 &  h=2 & h=3 & h=4 & h=6 & h=8 & h=10 & h=12 \\ \hline 
& DI &0.824 & 0.679 & 0.517 & 0.293 & -0.126 & -0.365 & -0.201 & -0.125 \\
& AR(p)  &0.838 & 0.682 & 0.502 & 0.245 & -0.231 & -0.435 & -0.263 & -0.125 \\
&PCA         & 0.516 & 0.393 & 0.300 & 0.231 & 0.149 & 0.089  & 0.030  & 0.011 \\
&Sq-PC  & 0.398 & 0.348 & 0.297 & 0.238 & 0.099 & -0.022 & -0.083 & -0.065 \\
&PC-Sq  & 0.605 & 0.488 & 0.391 & 0.296 & 0.186 & 0.090  & 0.017  & 0.044 \\
&kPCA  & 0.479 & 0.390 & 0.317 & 0.272 & 0.196 & 0.030  & -0.016 & -0.013 \\
& AE & 0.325 & 0.291 & 0.186 & -0.011 & 0.042 & 0.056 & 0.079 & 0.021 \\
&3PRF        & 0.597 & 0.484 & 0.429 & 0.369 & 0.273 & 0.111  & 0.083  & 0.176 \\
&k3PRF &0.760 & 0.640 & 0.478 & 0.605 & 0.433 & 0.199 & 0.169 & 0.389\\   \hline\hline

    \end{tabular}
    }
      \caption{$h$-period ahead out of sample $R^2$ of Macro Variables : Group-I}
      \label{table_macro_group1}
\end{table}

\begin{table}[!h]
    \centering
    \resizebox{130mm}{!}{
    \begin{tabular}{llccccccccc}
    \hline \\[1mm]
\textbf{Exports}& & & & & & &  & & \\  
\hline 
&   Method & h=1 &  h=2 & h=3 & h=4 & h=6 & h=8 & h=10 & h=12 \\ \hline 
& DI & 0.912 & 0.752 & 0.559 & 0.332 & -0.138 & -0.613 & -0.807 & -0.702  \\
&AR(p) &0.912 & 0.752 & 0.558 & 0.332 & -0.134 & -0.603 & -0.574 & -0.459\\
& PCA     & 0.353 & 0.306 & 0.248 & 0.193 & 0.123 & 0.107 & 0.106 & 0.109 \\
& Sq-PC  & 0.275 & 0.249 & 0.215 & 0.183 & 0.120 & 0.056 & 0.008 & -0.013 \\
& PC-Sq  & 0.399 & 0.326 & 0.243 & 0.166 & 0.073 & 0.066 & 0.113 & 0.194 \\
& kPCA  & 0.027 & 0.033 & 0.033 & 0.270 & 0.142 & -0.002 & -0.044 & 0.130 \\
& AE & 0.246 & 0.217 & 0.208 & 0.096 & 0.064 & 0.100 & 0.098 & -0.020 \\
& 3PRF        & 0.535 & 0.523 & 0.459 & 0.389 & 0.223 & 0.137 & 0.109 & 0.092 \\
& k3PRF & 0.724 & 0.705 & 0.641 & 0.602 & 0.546 & 0.575 & 0.600 & 0.631 \\[1mm]

\textbf{Imports}& & & & & & &  & & \\  
\hline
&    Method & h=1 &  h=2 & h=3 & h=4 & h=6 & h=8 & h=10 & h=12 \\ \hline 
& DI & 0.953 & 0.850 & 0.705 & 0.541 & 0.227 & -0.020 & 0.114 & 0.216  \\
&AR(p)       &0.952 & 0.842 & 0.689 & 0.518 & 0.191 & 0.056 & 0.169 & 0.211\\
&PCA         & 0.417 & 0.380 & 0.343 & 0.306 & 0.233 & 0.154 & 0.072 & 0.006 \\
&Sq-PC  & 0.395 & 0.373 & 0.341 & 0.299 & 0.194 & 0.079 & -0.005 & -0.046 \\
&PC-Sq  & 0.477 & 0.462 & 0.438 & 0.398 & 0.306 & 0.182 & 0.060 & 0.000 \\
&kPCA  & 0.421 & 0.389 & 0.348 & 0.311 & 0.241 & 0.081 & 0.064 & 0.033 \\
& AE & 0.399 & 0.243 & 0.255 & 0.418 & 0.276 & 0.080 & 0.093 & 0.042 \\
&3PRF        & 0.546 & 0.506 & 0.468 & 0.436 & 0.394 & 0.347 & 0.322 & 0.338 \\
&k3PRF &0.777 & 0.783 & 0.790 & 0.786 & 0.749 & 0.411 & 0.388 & 0.558\\[1mm]

\textbf{Fixed Invest.}& & & & & & &  & & \\   
\hline

&    Method & h=1 &  h=2 & h=3 & h=4 & h=6 & h=8 & h=10 & h=12 \\ \hline 
& DI &0.886 & 0.734 & 0.538 & 0.328 & -0.131 & -0.280 & -0.141 & -0.079\\
&AR(p)  &0.893 & 0.724 & 0.505 & 0.265 & -0.248 & -0.347 & -0.206 & -0.067\\
&PCA         & 0.490 & 0.384 & 0.290 & 0.220 & 0.134 & 0.088  & 0.042  & 0.016 \\
&Sq-PC  & 0.401 & 0.352 & 0.293 & 0.231 & 0.095 & -0.024 & -0.077 & -0.064 \\
&PC-Sq  & 0.595 & 0.492 & 0.385 & 0.314 & 0.208 & 0.104  & 0.030  & 0.068 \\
&kPCA  & 0.498 & 0.407 & 0.315 & 0.250 & 0.167 & 0.039  & -0.034 & 0.007 \\
& AE & 0.295 & 0.235 & 0.289 & 0.229 & 0.142 & 0.145 & 0.101 & -0.094 \\
&3PRF        & 0.525 & 0.454 & 0.389 & 0.348 & 0.251 & 0.122  & 0.127  & 0.226 \\
&k3PRF &0.736 & 0.659 & 0.426 & 0.578 & 0.265 & 0.235 & 0.261 & 0.359\\[1mm]

\textbf{IP : Final }& & & & & & &  & & \\  
\hline
 &   Method & h=1 &  h=2 & h=3 & h=4 & h=6 & h=8 & h=10 & h=12 \\ \hline 
& DI &0.943 & 0.866 & 0.764 & 0.647 & 0.376 & 0.080 & -0.187 & -0.461 \\
& AR(p) &0.952 & 0.871 & 0.765 & 0.644 & 0.362 & 0.059 & -0.222 & -0.562\\
& PCA & 0.632 & 0.541 & 0.439 & 0.331 & 0.135 & -0.021 & -0.055 & -0.041 \\
& Sq-PC & 0.530 & 0.483 & 0.421 & 0.344 & 0.167 & 0.010 & -0.060 & -0.073 \\
& PC-Sq & 0.701 & 0.615 & 0.505 & 0.381 & 0.162 & 0.012 & -0.051 & -0.030 \\
& kPCA & 0.637 & 0.570 & 0.505 & 0.456 & 0.340 & 0.036 & 0.047 & 0.012 \\
& AE & 0.406 & 0.558 & 0.367 & 0.268 & 0.141 & 0.068 & -0.150 & -0.043 \\
& 3PRF & 0.677 & 0.640 & 0.579 & 0.487 & 0.348 & 0.192 & 0.134 & 0.225 \\
& k3PRF & 0.828 & 0.778 & 0.731 & 0.700 & 0.499 & 0.263 & 0.283 & 0.731 \\
\hline \hline 
\end{tabular}
    }
      \caption{$h$-period ahead out of sample $R^2$ of Macro Variables : Group-II}
  \label{table_macro_group2}
\end{table}

To present the results in an organized manner, we create two tables. In Table-\ref{table_macro_group1}, we display the forecasting performance for three series: GDP, Consumption, and Investment, which we informally refer to as `Group-I'. Table-\ref{table_macro_group2} presents a comparative analysis of forecasting performance for `Group-II' macro variables\footnote{ Variables' FRED-QD code and description can be found in Online Appendix-B.4}: Exports, Imports, Fixed Investments, and Industrial Production (Final Index). As defined earlier in the text, the reported numbers in the tables represent out-of-sample $R^2$ values across various forecast horizons ranging from one period ahead to twelve periods ahead.\\[2mm]
Results highlight a secular observation that among various unsupervised forecasting methodologies—PC, Squared-PC, PC-Squared, and non-linear unsupervised approaches such as kernel PCA—none exhibit superior performance compared to our proposed method across any forecast horizon for the seven series under consideration. While the supervised linear forecasting model 3PRF demonstrates improved performance relative to the unsupervised techniques, it still falls short of our non-linear supervised approach. Notably, the autoregressive (AR) model and its variant, the Diffusion Index (DI), emerge as the sole contenders capable of surpassing our method at shorter horizons, albeit only marginally and for a few series. The neural-network-based competitor, the Autoencoder (AE), does not outperform our method at any forecast horizon—a result that is not surprising, as neural-network-based methods are known to require large sample sizes for effective training, which is not the case here.

In summary, our method significantly outperforms all competitors across longer horizons and competes well with AR(p) and DI in the short run.  Therefore, it serves as a dependable and preferred forecasting framework across all forecast horizons in macroeconomic prediction tasks.

\FloatBarrier
\subsection{Forecasting Labor Market and Price Variables}
This analysis aims to forecast key labor market and price variables. Within the labor market category, we focus on unemployment rates and total non-farm employment (Nonfarm Emp). We examine the GDP Deflator and the Consumer Price Index (CPI) for price variables. The GDP Deflator offers insights into overall inflation at the macroeconomic level, while the CPI captures inflation experienced by consumers at a more disaggregated level. The results of this analysis are summarized in Table \ref{table_labor_forecast}.

\begin{table}[!h]
    \centering
     \resizebox{130mm}{!}{
   
    \begin{tabular}{llccccccccc}
  \hline
\textbf{Nonfarm Emp}  & & & & & & &  & &\\
    \hline
&   Method & h=1 &  h=2 & h=3 & h=4 & h=6 & h=8 & h=10 & h=12 \\ \hline 
& DI &0.971 & 0.911 & 0.798 & 0.641 & 0.226 & -0.262 & -0.723 & -0.843  \\
&AR(p)&0.967 & 0.878 & 0.737 & 0.555 & 0.098 & -0.410 & -0.861 & -0.991\\
&PCA         & 0.786 & 0.728 & 0.604 & 0.435 & 0.057 & -0.219 & -0.258 & -0.146 \\
&Sq-PC  & 0.528 & 0.498 & 0.440 & 0.361 & 0.167 & -0.024 & -0.109 & -0.098 \\
&PC-Sq  & 0.836 & 0.795 & 0.679 & 0.510 & 0.131 & -0.139 & -0.210 & -0.110 \\
&kPCA  & 0.832 & 0.790 & 0.702 & 0.587 & 0.370 & 0.196 & 0.112 & 0.059 \\
& AE & 0.503 & 0.460 & 0.340 & 0.279 & 0.014 & -0.026 & -0.256 & -0.021 \\
&3PRF        & 0.765 & 0.731 & 0.712 & 0.662 & 0.407 & 0.312 & 0.264 & 0.229 \\
&k3PRF &0.929 & 0.895 & 0.846 & 0.768 & 0.556 & 0.444 & 0.441 & 0.584\\[2mm]

\textbf{Unemp Rate}  & & & & & & &  & &\\
    \hline
&     Method & h=1 &  h=2 & h=3 & h=4 & h=6 & h=8 & h=10 & h=12 \\ \hline 
& DI & 0.962 & 0.921 & 0.856 & 0.774 & 0.576 & 0.423 & 0.344 & 0.292 \\
&AR(p) &0.962 & 0.905 & 0.828 & 0.737 & 0.541 & 0.323 & 0.179 & 0.107\\
&PCA         & 0.810 & 0.853 & 0.849 & 0.809 & 0.648 & 0.426 & 0.255 & 0.133 \\
&Sq-PC  & 0.825 & 0.852 & 0.849 & 0.821 & 0.686 & 0.457 & 0.251 & 0.097 \\
&PC-Sq  & 0.798 & 0.849 & 0.851 & 0.820 & 0.687 & 0.497 & 0.304 & 0.225 \\
&kPCA  & 0.610 & 0.664 & 0.672 & 0.675 & 0.647 & 0.562 & 0.440 & -0.035 \\
& AE & 0.462 & 0.583 & 0.485 & 0.508 & 0.326 & 0.349 & 0.005 & 0.008 \\
&3PRF        & 0.913 & 0.914 & 0.863 & 0.802 & 0.638 & 0.475 & 0.402 & 0.471 \\
&k3PRF &0.924 & 0.937 & 0.903 & 0.846 & 0.674 & 0.508 & 0.459 & 0.390 \\ 
\hline \\[2mm]

 \textbf{GDP Deflator}  & & & & & & &  & &\\
  \hline
&     Method & h=1 &  h=2 & h=3 & h=4 & h=6 & h=8 & h=10 & h=12 \\ \hline 
&DI& 0.762 & 0.572 & 0.392 & 0.194 & 0.124 & 0.090 & 0.121 & 0.104 \\
&AR(p) &0.794 & 0.589 & 0.422 & 0.253 & 0.276 & 0.278 & 0.294 & 0.263\\
&PCA   & 0.444 & 0.276 & 0.056 & -0.184 & -0.408 & -0.347 & -0.221 & -0.057 \\
&Sq-PC  & 0.299 & 0.145 & -0.035 & -0.168 & -0.245 & -0.230 & -0.192 & -0.108 \\
&PC-Sq  & 0.431 & 0.268 & 0.104 & -0.039 & -0.106 & -0.038 & -0.111 & -0.182 \\
&kPCA  & -0.032 & 0.247 & -0.021 & 0.008 & 0.003 & 0.004 & 0.029 & -0.023 \\
&AE& 0.045 & -0.014 & 0.256 & -0.172 & -0.366 & -0.202 & -0.142 & 0.061 \\
&3PRF        & 0.584 & 0.496 & 0.426 & 0.243 & 0.174 & 0.279 & 0.300 & 0.155 \\
&k3PRF &0.667 & 0.632 & 0.563 & 0.476 & 0.479 & 0.413 & 0.197 & 0.512\\[1mm]
\textbf{CPI}  & & & & & & &  & &\\
\hline
&     Method & h=1 &  h=2 & h=3 & h=4 & h=6 & h=8 & h=10 & h=12 \\ \hline 
& DI &0.753 & 0.617 & 0.436 & 0.341 & 0.401 & 0.411 & 0.448 & 0.470\\
&AR(p) &0.718 & 0.591 & 0.395 & 0.193 & 0.396 & 0.494 & 0.487 & 0.519\\
&PCA         & 0.660 & 0.535 & 0.364 & 0.154 & -0.163 & -0.252 & -0.248 & -0.173 \\
&Sq-PC  & 0.410 & 0.296 & 0.161 & 0.049 & -0.055 & -0.156 & -0.200 & -0.173 \\
&PC-Sq  & 0.649 & 0.512 & 0.353 & 0.186 & -0.019 & -0.087 & -0.187 & -0.228 \\
&kPCA  & 0.440 & 0.380 & -0.050 & 0.189 & -0.043 & -0.024 & 0.042 & -0.006 \\
& AE & 0.315 & 0.195 & 0.066 & 0.053 & 0.132 & -0.091 & -0.223 & -0.037 \\
&3PRF        & 0.641 & 0.566 & 0.487 & 0.352 & 0.192 & 0.241 & 0.255 & 0.141 \\
&k3PRF &0.676 & 0.612 & 0.541 & 0.463 & 0.469 & 0.434 & 0.349 & 0.477\\
\hline  \hline
     \end{tabular}
    }
     \caption{Out of Sample $R^2$ of Labor Market and Price Variables}
       \label{table_labor_forecast}
\end{table}
\noindent While the major story remains of the forecast performance results are qualitatively similar to those of aggregate macroeconomic series forecasting. We want to highlight the following: in the Unemployment Rate and GDP Deflator, our method beats AR(p) and DI method even in the short-run, barring $h=1$, and in the remaining two variables, our method is the best except initial two horizons.

\FloatBarrier
\subsection{Forecasting Housing and Financial Variables}
We evaluate the relative performance of our method across several key indicators: Privately Owned Housing Starts (\textit{HStart}), Privately Owned Housing Starts in the Western Census region (\textit{HStart-W}), GS-1 (Treasury Bills), GS-10 (Treasury Notes), and the S\&P 500 Index. The first two indicators pertain to the housing market, while the latter three belong to the financial market. These financial variables are listed in ascending order of volatility.
\begin{table}[!h]
    \centering
    \resizebox{130mm}{!}{
    \begin{tabular}{llccccccccc}
  \hline 
    \textbf{HStart}  & & & & & & &  & &\\
\hline
    
    & Method & h=1 &  h=2 & h=3 & h=4 & h=6 & h=8 & h=10 & h=12 \\ \hline 
& DI &0.400 & 0.533 & 0.610 & 0.637 & 0.401 & 0.358 & 0.284 & 0.314\\
&AR(p) &0.362 & 0.437 & 0.556 & 0.558 & 0.334 & 0.390 & 0.422 & 0.342 \\
&PCA & -1.360 & -0.799 & -0.317 & -0.052 & 0.172 & 0.259 & 0.086 & 0.085 \\
&Sq-PC & -1.226 & -0.688 & -0.196 & 0.095 & 0.314 & 0.453 & 0.183 & 0.100 \\
&PC-Sq & -1.473 & -0.936 & -0.371 & -0.004 & 0.278 & 0.188 & -0.176 & -0.024 \\
&kPCA & -0.199 & -0.074 & -0.157 & 0.244 & 0.408 & -0.101 & -0.325 & 0.101 \\
& AE & -1.517 & -0.663 & -0.220 & 0.272 & 0.035 & -0.412 & -0.460 & -0.242 \\
&3PRF & 0.092 & 0.272 & 0.064 & -0.223 & -0.391 & -0.205 & -0.220 & -0.653 \\
&k3PRF & 0.138 & 0.204 & 0.231 & 0.245 & 0.230 & 0.253 & 0.116 & 0.073 \\[2mm]
\textbf{HStart-W}  & & & & & & &  & &\\
\hline 
&     Method & h=1 &  h=2 & h=3 & h=4 & h=6 & h=8 & h=10 & h=12 \\ \hline 
& DI &0.645 & 0.636 & 0.526 & 0.449 & 0.400 & 0.228 & 0.161 & 0.159\\
&AR(p) &0.585 & 0.562 & 0.411 & 0.287 & 0.223 & 0.032 & -0.037 & 0.045 \\
&PCA & 0.326 & 0.433 & 0.481 & 0.516 & 0.405 & 0.169 & -0.070 & -0.182 \\
&Sq-PC & 0.201 & 0.318 & 0.356 & 0.372 & 0.184 & -0.053 & -0.248 & -0.323 \\
&PC-Sq & 0.359 & 0.402 & 0.414 & 0.459 & 0.310 & 0.033 & -0.135 & -0.244 \\
&kPCA & 0.287 & 0.336 & 0.379 & 0.442 & 0.447 & -0.135 & -0.147 & -0.062 \\
& AE & -0.336 & 0.168 & 0.215 & 0.191 & 0.133 & 0.021 & -0.089 & -0.271 \\
&3PRF & 0.571 & 0.475 & 0.231 & 0.084 & -0.031 & 0.094 & 0.260 & 0.253 \\
&k3PRF & 0.586 & 0.464 & 0.207 & 0.554 & 0.178 & 0.141 & 0.160 & 0.463 \\[1mm]
\hline \hline 
    \end{tabular}
    }
      \caption{Out of Sample $R^2$ of Housing and Financial Variables}
  
    \label{table_housing_forecast}
\end{table}

\begin{table}[!h]
    \centering
    \resizebox{130mm}{!}{
    \begin{tabular}{llccccccccc}
  \hline 
    \textbf{GS-1}  & & & & & & &  & &\\
    \hline
&     Method & h=1 &  h=2 & h=3 & h=4 & h=6 & h=8 & h=10 & h=12 \\ \hline 
& DI &0.759 & 0.553 & 0.304 & 0.010 & -0.380 & -0.596 & -0.644 & -0.640\\
&AR(p)       &0.901 & 0.751 & 0.558 & 0.297 & -0.172 & -0.472 & -0.114 & -0.033\\
&PCA         & 0.687 & 0.487 & 0.261 & 0.055 & -0.163 & -0.124 & -0.033 & 0.139 \\
&Sq-PC  & 0.306 & 0.201 & 0.090 & -0.012 & -0.145 & -0.131 & -0.074 & 0.011 \\
&PC-Sq  & 0.674 & 0.448 & 0.243 & 0.059 & -0.162 & -0.119 & 0.051 & 0.163 \\
&kPCA  & 0.635 & 0.472 & 0.282 & 0.119 & 0.029 & -0.018 & 0.166 & 0.114 \\
& AE & 0.277 & 0.057 & 0.120 & -0.021 & -0.193 & 0.136 & 0.219 & 0.220 \\
&3PRF        & 0.856 & 0.735 & 0.615 & 0.501 & 0.449 & 0.329 & 0.241 & 0.349 \\
&k3PRF &0.873 & 0.806 & 0.782 & 0.699 & 0.381 & 0.224 & 0.428 & 0.605\\
\hline \\[1mm]
    \textbf{GS-10}  & & & & & & &  & &\\
    \hline
&       Method & h=1 &  h=2 & h=3 & h=4 & h=6 & h=8 & h=10 & h=12 \\ \hline 
& DI & 0.694 & 0.502 & 0.327 & 0.028 & -0.078 & -0.013 & 0.203 & 0.392\\
&AR(p) &0.791 & 0.613 & 0.480 & 0.183 & -0.093 & 0.226 & 0.425 & 0.430\\
&PCA         & 0.446 & 0.327 & 0.148 & 0.017 & -0.122 & -0.177 & -0.329 & -0.378 \\
&Sq-PC  & 0.312 & 0.247 & 0.124 & 0.069 & -0.016 & -0.065 & -0.194 & -0.327 \\
&PC-Sq  & 0.421 & 0.292 & 0.200 & 0.155 & 0.117 & 0.032 & -0.083 & -0.608 \\
&kPCA  & 0.457 & 0.402 & -0.098 & 0.246 & -0.039 & -0.022 & 0.082 & 0.035 \\
& AE & 0.158 & 0.263 & -0.259 & 0.019 & 0.006 & -0.260 & 0.036 & -0.168 \\
&3PRF        & 0.615 & 0.469 & 0.268 & 0.012 & 0.168 & 0.403 & 0.294 & 0.044 \\
&k3PRF &0.621 & 0.499 & 0.405 & 0.401 & 0.345 & 0.272 & 0.161 & 0.566\\
\hline\\[1mm]

    \textbf{S\&P 500}  & & & & & & &  & &\\
    \hline
&       Method & h=1 &  h=2 & h=3 & h=4 & h=6 & h=8 & h=10 & h=12 \\ \hline 
& DI &0.932 & 0.802 & 0.641 & 0.463 & 0.146 & -0.004 & -0.002 & -0.013\\
&AR(p)       &0.934 & 0.805 & 0.649 & 0.480 & 0.184 & 0.054 & 0.064 & 0.024 \\
&PCA         & 0.388 & 0.318 & 0.224 & 0.121 & -0.019 & -0.001 & 0.107 & 0.201 \\
&Sq-PC  & 0.265 & 0.214 & 0.152 & 0.089 & 0.023 & 0.061 & 0.136 & 0.192 \\
&PC-Sq  & 0.387 & 0.287 & 0.167 & 0.048 & -0.079 & 0.034 & 0.220 & 0.295 \\
&kPCA  & -0.064 & -0.067 & -0.039 & -0.031 & 0.094 & 0.038 & 0.091 & 0.558 \\
& AE & 0.359 & 0.273 & 0.270 & 0.267 & 0.091 & 0.199 & 0.094 & 0.280 \\
&3PRF        & 0.706 & 0.687 & 0.636 & 0.566 & 0.453 & 0.458 & 0.489 & 0.523 \\
&k3PRF &0.812 & 0.791 & 0.736 & 0.654 & 0.565 & 0.586 & 0.674 & 0.781\\
\hline \hline

    \end{tabular}
    }
      \caption{Out of Sample $R^2$ of Housing and Financial Variables}
  
    \label{table_fin_forecast}
\end{table}

\noindent We observe that the linear AR(p) and DI models perform exceptionally well for housing variables, with our method closely trailing them while outperforming all other competitors. In contrast, for financial variables, our method decisively surpasses all competing approaches as the forecast horizon extends to $h \geq 3$.

\FloatBarrier
\subsection{Robustness Check: Evaluating Forecasts After Partialing Out AR(4) Variation}
We observe that the AR(p) model and its closely related method, DI, perform particularly well in the short run. It is well-established in the literature that AR(p) models are difficult to outperform over shorter forecast horizons. Possibly for this reason, some forecasting approaches evaluate performance after accounting for the variation captured by AR(p). For instance, our closest competitor \cite{kelly} effectively forecasts the innovation of an AR(4) process for each variable, thereby “partialing out” the predictive information contained in the target variable’s own lags.

Following a similar approach, we also conduct an exercise in which we partial out AR(4) variation and subsequently compare the forecasting performance of all methods. We find that the qualitative patterns remain consistent, and in some cases are even more pronounced. The detailed results are presented in Online Appendix B.8.

\section{Comprehensive Forecasting Analysis}
To enhance the robustness of our empirical analysis, we conducted comparative assessments of our method against competing methods across all 176 series within our dataset. This entailed selecting each series as the target and repeating the comparative analysis for every series in our dataset. 

\subsection{Description of Comparisons}
Our investigation encompasses the comparative performance of models across a total of $176\times 8 =1408$ target-horizon combinations. The results of these comparisons indicate the percentage of instances where a particular method demonstrated superior performance to all other competitors. For example, if a method emerged as the best performer in 704 out of 1408 combinations, it would be represented by a value of 50 in the table. Essentially, we list the relative frequency of the occurrence of the best performance of a given method.

While the preceding frequency comparisons provide insight into the number of times each method proved superior to others, they do not measure the extent to which the best-performing method surpassed its nearest competitor. In other words, while method A may marginally outperform method B on one forecast horizon, method B might exhibit a considerable advantage over method A on another horizon. Then, the aforementioned frequency comparison may not depict the full picture. To account for this, we introduce a notion of `\textit{Tolerance}' level. We call a method `best' under tolerance level $\epsilon$ if the out-of-sample $R^2$ of a method is within $\epsilon$ percentage lower than the best method's performance \footnote{For example, if the AR model is the best for a of series $y_\ell$ and horizon $h_0$ with a $R^2=0.60$. For tolerance=5, another method will also be considered `best' if its $R^2 \geq 0.60(1-5/100)=0.57$}. Therefore, for a non-zero tolerance, it is possible to have multiple `best' methods. \\[2mm]
The ``All Horizons'' set of rows summarizes all 1408 comparisons, i.e., encompassing all horizons and all series. Recognizing that forecast objectives may vary in time horizon, we scrutinize comparative performances in short- and long-run contexts. The ``Short-run" rows incorporate horizons $h={1,2,3,4}$, comprising 708 (calculated as $176 \times 4$) combinations, while the ``Long-run" row includes horizons $h={6,8,10,12}$, similarly amounting to 708 combinations. Additionally, the portion labeled as ``Excluding AR" excludes the auto-regressive AR(p) and closely related Diffusion Index (DI) methods and compares the remaining methods across all 1408 combinations. 
For more granular analysis, we report comparative performance numbers for each forecast horizon $h$. These numbers are reported in the Online Appendix-B.7.   

It is important to note that multiple `best' methods may exist for a non-zero tolerance level, resulting in the sums of rows (in Table 5 in Online Appendix-B.7) exceeding 100 percent. However, for a tolerance level of zero, the rows sum to 100 percent.

\FloatBarrier
\subsection{Results}
\begin{table}[!h]
    \centering
\resizebox{160mm}{!}{
    \begin{tabular}{lcccccccccc}
    \hline\hline
    
    \multicolumn{1}{c}{\textbf{Analysis}}& \multicolumn{1}{c}{\textbf{Tolerance(\%)}} & \multicolumn{7}{c}{\textbf{Methods}} \\
    \hline

 & & \textbf{DI} &\textbf{AR(p)}  & \textbf{PCA} & \textbf{Sq-PC} & \textbf{PC-Sq} & \textbf{kPCA}& \textbf{AE} & \textbf{3PRF} & \textbf{k3PRF} \\
    \hline
    \textbf{Overall} & & &  & &   &   &  &&  &  \\
  
    &\textbf{0}  &15.63 & 18.54 & 0.28 & 0.92 & 1.78 & 3.69 & 0.28 & 7.74 & 51.14\\
    &\textbf{5}   &28.69 & 30.82 & 1.21 & 1.42 & 2.27 & 4.19 & 0.36 & 10.80 & 56.39\\
    &\textbf{10} &35.16 & 36.72 & 2.63 & 2.34 & 3.98 & 5.18 & 0.43 & 17.19 & 62.00 \\
    &\textbf{20} &42.26 & 43.75 & 6.82 & 3.98 & 9.94 & 8.38 & 0.64 & 30.47 & 73.51\\[2mm]
    \textbf{Short-run} & & &  & &   &   &  & \\
    &\textbf{0} &26.42 & 29.97 & 0.43 & 0.57 & 1.70 & 1.70 & 0.14 & 3.98 & 35.09 \\
    &\textbf{5} &50.28 & 51.42 & 1.70 & 1.14 & 2.70 & 2.27 & 0.14 & 8.38 & 43.61\\
    &\textbf{10} &60.51 & 61.79 & 3.98 & 2.41 & 5.68 & 2.98 & 0.14 & 17.19 & 51.85\\
    &\textbf{20} &70.88 & 71.16 & 10.94 & 4.26 & 15.20 & 6.39 & 0.14 & 33.24 & 69.18\\[2mm]
   
    \textbf{Long-run} & & &  & &   &   &  & \\
    &\textbf{0}   &4.83 & 7.10 & 0.14 & 1.28 & 1.85 & 5.68 & 0.43 & 11.51 & 67.19 \\
    &\textbf{5} &7.10 & 10.23 & 0.71 & 1.70 & 1.85 & 6.11 & 0.57 & 13.21 & 69.18\\
    &\textbf{10} &9.80 & 11.65 & 1.28 & 2.27 & 2.27 & 7.39 & 0.71 & 17.19 & 72.16\\
    &\textbf{20} &13.64 & 16.34 & 2.70 & 3.69 & 4.69 & 10.37 & 1.14 & 27.70 & 77.84 \\[2mm]
    
    \textbf{Without AR} & & &  & &   &   &  & \\
    &\textbf{0}   &- & - & 1.21 & 1.42 & 2.84 & 5.47 & 0.71 & 12.86 & 75.50\\
    &\textbf{5} &- & - & 2.70 & 1.99 & 4.62 & 5.68 & 0.85 & 17.83 & 78.55\\
    &\textbf{10} &- & - & 5.11 & 3.20 & 7.53 & 6.96 & 1.07 & 25.99 & 81.39\\
    &\textbf{20} &- & - & 11.01 & 5.75 & 14.20 & 11.43 & 1.49 & 41.34 & 85.94\\
    \hline \hline 
    \end{tabular}
    }
    \caption{Distribution of Best Forecasting Methods Across All Series in Our Data (Percentage)}
      \small{Notes: We call a method the \textit{best} if either it offers maximum out of sample (OOS) $R^2$ or its OOS $R^2$ falls within a tolerance percentage of the highest OOS $R^2$ for a given row. We run this experiment on all 176 quarterly time series available in the FRED-QD dataset for 8 forecast horizons i.e. a total of $176 \times 8 =1408$ comparisons.}

    \label{table_all_series_main}
\end{table}

We present the results in table-\ref{table_all_series_main}.
The findings presented above yield several noteworthy observations. First, it is evident that unsupervised forecasting techniques, including PC, Squared-PC, PC-squared, Autoencoder, and kernel PCA, exhibit inferior performance across the majority of scenarios when compared to our method. Second, our method, kernel 3PRF, demonstrates unequivocal superiority in longer-horizon forecasting endeavors. Third, our method is unequivocally superior across all horizons when autoregressive (AR) and closely related DI methods are excluded. Fourth, even in the short run, our method is the single best-performing approach under strict (zero-tolerance) evaluations of absolute performance. Moreover, its performance remains comparable to the leading short-run benchmarks, AR(p) and DI, when we allow for some tolerance level.

Taken together, by demonstrating consistent performance across all series in our dataset, we conclude that our method can be regarded as a reliable forecasting tool across all forecast horizons.

\FloatBarrier
\section{Conclusion}
Building upon the three-pass regression filter by \cite{kelly},  we introduce a new forecasting method, kernel three-pass regression filter. 
Through extensive empirical exercises, we show that this approach holds promise as a dependable forecasting tool. Improved performance can be attributed to two noteworthy features of our method. First, it integrates non-linear relationships by transforming input data into a higher-dimensional space, encapsulating its non-linear functions. Second, it operates as a supervised method, effectively filtering out and discarding irrelevant factors while predicting the target variable.





\bibliographystyle{authordate1}
 \bibliography{main.bib}

\appendix

\section{Technical Appendix } \label{apx_proofs}
\subsection{Proofs of Theoretical Results} \label{apx_proofs}

\begin{lem} \label{l1}
    
\noindent Under Assumption(s) \ref{asu1}-\ref{asu3}, we have the following
\begin{enumerate}
    \item   $T^{-1 / 2} \boldsymbol{F}^{\prime} \boldsymbol{J}_{T} \boldsymbol{\omega}=\boldsymbol{O}_{p}(1)$ \label{2.1}

    \item    $T^{-1 / 2} \boldsymbol{F}^{\prime} \boldsymbol{J}_{T} \boldsymbol{\eta}=\boldsymbol{O}_{p}(1)$ \label{2.2}

    \item    $T^{-1 / 2} \boldsymbol{\varepsilon}^{\prime} \boldsymbol{J}_{T} \boldsymbol{\eta}=\boldsymbol{O}_{p}(1)$ \label{2.3}

    \item   $M^{-1 / 2} \varepsilon_{t}^{\prime}  \boldsymbol{\Phi}=\boldsymbol{O}_{p}(1)$ \label{2.4}

    \item  $M^{-1} T^{-1} \boldsymbol{\Phi}^{\prime}  \boldsymbol{\varepsilon}^{\prime} \boldsymbol{J}_{T} \boldsymbol{F}=\boldsymbol{O}_{p}\left(\delta_{M T}^{-1}\right)$  \label{2.5}

    \item   $M^{-1} T^{-1/2 } \boldsymbol{\Phi}^{\prime}  \boldsymbol{\varepsilon}^{\prime} \boldsymbol{J}_{T} \boldsymbol{\omega}=\boldsymbol{O}_{p}\left(1\right)$ \label{l2.6}

    \item   $M^{-1 / 2} T^{-1/2 } \boldsymbol{\Phi}  \boldsymbol{\varepsilon}^{\prime} \boldsymbol{J}_{T} \boldsymbol{\eta}=\boldsymbol{O}_{p}(1)$ \label{2.7}

    \item    $M^{-1} T^{-3 / 2} \boldsymbol{F}^{\prime} \boldsymbol{J}_{T} \boldsymbol{\varepsilon}  \boldsymbol{\varepsilon}^{\prime} \boldsymbol{J}_{T} \boldsymbol{F}=\boldsymbol{O}_{p}\left(\delta_{M T}^{-1}\right)$ \label{2.8}

    \item    $M^{-1} T^{-3 / 2} \boldsymbol{\omega}^{\prime} \boldsymbol{J}_{T} \varepsilon  \boldsymbol{\varepsilon}^{\prime} \boldsymbol{J}_{T} \boldsymbol{F}=\boldsymbol{O}_{p}\left(\delta_{M T}^{-1}\right)$ \label{2.9}

    \item   $M^{-1} T^{-3 / 2} \boldsymbol{\omega}^{\prime} \boldsymbol{J}_{T} \varepsilon  \varepsilon^{\prime} \boldsymbol{J}_{T} \boldsymbol{\omega}=\boldsymbol{O}_{p}\left(\delta_{M T}^{-1}\right)$ \label{2.10}

    \item    $M^{-1} T^{-1 / 2} \boldsymbol{F}^{\prime} \boldsymbol{J}_{T} \varepsilon  \boldsymbol{\varepsilon}_{t}=\boldsymbol{O}_{p}\left(\delta_{M T}^{-1}\right)$ \label{2.11}

    \item    $M^{-1} T^{-1 / 2} \boldsymbol{\omega}^{\prime} \boldsymbol{J}_{T} \varepsilon  \varepsilon_{t}=\boldsymbol{O}_{p}\left(\delta_{M T}^{-1}\right)$ \label{2.12}

    \item   $M^{-1} T^{-3 / 2} \boldsymbol{\eta}^{\prime} \boldsymbol{J}_{T} \varepsilon  \varepsilon^{\prime} \boldsymbol{J}_{T} \boldsymbol{F}=\boldsymbol{O}_{p}\left(\delta_{M T}^{-1}\right)$ \label{2.13}

    \item   $M^{-1} T^{-3 / 2} \boldsymbol{\eta}^{\prime} \boldsymbol{J}_{T} \varepsilon  \varepsilon^{\prime} \boldsymbol{J}_{T} \boldsymbol{F}=\boldsymbol{O}_{p}\left(\delta_{M T}^{-1}\right)$ \label{2.14}

     \item $T^{-1 / 2} \sum_{t} \eta_{t+h}=O_{p}(1)$ \label{2.15}
\end{enumerate}
\end{lem}

\noindent \textit{Proof:} The result follows from \cite{kelly}, Lemma 2 in their appendix. In their setup, the matrix $\boldsymbol{J}_N$ appears in these expressions, whereas in our case there is no analogous $\boldsymbol{J}_M$. Consequently, if we expand the matrix products as in \cite{kelly}, fewer terms arise compared to their setting. The order of the remaining terms was established in \cite{kelly}, with the only adjustment here being that we work with $\varphi(\boldsymbol{X})$ instead of $\boldsymbol{X}$, so the feature dimension $M$ appears in the relevant norms and sums. Therefore, the convergence rates of these matrix products are at least as fast, if not faster, than analogous products in \cite{kelly}.

 Assumptions \ref{asu3}.\ref{i3.5}-\ref{asu3}.\ref{i3.7} in our paper effectively replace Assumption 4 of \cite{kelly}, which imposes a distributional structure on these stochastic processes. For the results featuring in Lemma 1 (subsequently used in Lemma 2) of \cite{kelly}, they only use the boundedness of second moments of these processes, which is precisely what we assume in \ref{asu3}.\ref{i3.5}-\ref{asu3}.\ref{i3.7}. Therefore, the convergence results here remain valid. The stronger distributional assumptions in \cite{kelly} are used only to establish convergence in distribution of the factors, coefficients, and target, which we do not pursue here due to the concerns flagged in the remarks of our paper.

\begin{lem} \label{l2} \noindent Under Assumption(s) \ref{asu1}-\ref{asu5}, we have the following
  \begin{enumerate}
      \item $ M^{-1} T^{-1} \boldsymbol{Z}^{\prime} \boldsymbol{J}_{T} \varphi(\boldsymbol{X}) \varphi( \boldsymbol{x}_{t}) = \boldsymbol{\Lambda} \boldsymbol{\Delta}_{F} \mathcal{P} \boldsymbol{F}_{t} + \boldsymbol{O}_{p}\left(\delta_{M T}^{-1}\right)$ \label{l2.1}

      \item $M^{-1} T^{-2} \boldsymbol{Z}^{\prime} \boldsymbol{J}_{T} \varphi(\boldsymbol{X} )\varphi( \boldsymbol{X})^{\prime}\boldsymbol{J}_{T} \boldsymbol{y}$ $= \boldsymbol{\Lambda} \boldsymbol{\Delta}_{F} \mathcal{P} \boldsymbol{\Delta}_{F} \boldsymbol{\beta} + \boldsymbol{O}_{p}\left(\delta_{M T}^{-1}\right)$ \label{l2.2}

      \item $M^{-2} T^{-3} \boldsymbol{Z}^{\prime} \boldsymbol{J}_{T} \varphi(\boldsymbol{X} )\varphi( \boldsymbol{X})^{\prime} \boldsymbol{J}_{T} \varphi(\boldsymbol{X} )\varphi( \boldsymbol{X})^{\prime} \boldsymbol{J}_{T} \boldsymbol{Z}$ $=\boldsymbol{\Lambda} \boldsymbol{\Delta}_{F} \mathcal{P} \boldsymbol{\Delta}_{F} \mathcal{P} \boldsymbol{\Delta}_{F} \boldsymbol{\Lambda}^{\prime} + \boldsymbol{O}_{p}\left(\delta_{M T}^{-1}\right)$ \label{l2.3}
  \end{enumerate}  
\end{lem}

\noindent \textit{Proof:} The Proof follows directly by writing out the expressions. Item 1
\begin{align*}
 M^{-1} T^{-1} \boldsymbol{Z}^{\prime} \boldsymbol{J}_{T} \varphi(\boldsymbol{X}) \varphi( \boldsymbol{x}_{t}) &= 
\boldsymbol{\Lambda}\left(T^{-1} \boldsymbol{F}^{\prime} \boldsymbol{J}_{T} \boldsymbol{F}\right)\left(M^{-1} \boldsymbol{\Phi}^{\prime}  \boldsymbol{\Phi}\right) \boldsymbol{F}_{t} 
 +\boldsymbol{\Lambda}\left(T^{-1} \boldsymbol{F}^{\prime} \boldsymbol{J}_{T} \boldsymbol{F}\right)\left(M^{-1} \boldsymbol{\Phi}^{\prime} \boldsymbol{\varepsilon}_{t}\right) \\
& +\boldsymbol{\Lambda}\left(M^{-1} T^{-1} \boldsymbol{F}^{\prime} \boldsymbol{J}_{T} \boldsymbol{\varepsilon}  \boldsymbol{\Phi}\right) \boldsymbol{F}_{t} 
 +\boldsymbol{\Lambda}\left(M^{-1} T^{-1} \boldsymbol{F}^{\prime} \boldsymbol{J}_{T} \boldsymbol{\varepsilon} \boldsymbol{\varepsilon}_{t}\right) \\
&+\left(T^{-1} \boldsymbol{\omega}^{\prime} \boldsymbol{J}_{T} \boldsymbol{F}\right)\left(M^{-1} \boldsymbol{\Phi}^{\prime} \boldsymbol{\Phi}\right) \boldsymbol{F}_{t} 
 +\left(T^{-1} \boldsymbol{\omega}^{\prime} \boldsymbol{J}_{T} \boldsymbol{F}\right)\left(M^{-1} \boldsymbol{\Phi}^{\prime}  \boldsymbol{\varepsilon}_{t}\right) \\
&+\left(M^{-1} T^{-1} \boldsymbol{\omega}^{\prime} \boldsymbol{J}_{T} \boldsymbol{\varepsilon}  \boldsymbol{\Phi}\right) \boldsymbol{F}_{t} 
 +\left(M^{-1} T^{-1} \boldsymbol{\omega}^{\prime} \boldsymbol{J}_{T} \boldsymbol{\varepsilon}\boldsymbol{\varepsilon}_{t}\right) \\
&  =\boldsymbol{\Lambda} \boldsymbol{\Delta}_{F} \mathcal{P} \boldsymbol{F}_{t} + \boldsymbol{O}_{p}\left(\delta_{M T}^{-1}\right)
\end{align*}
 The final line follows directly from Lemma \ref{l1} and Assumptions \ref{asu2}.\ref{i2.1} and \ref{asu2}.\ref{i2.2}.

\vspace{4mm}
 \noindent  Item 2:
\begin{align*}
& M^{-1} T^{-2} \boldsymbol{Z}^{\prime} \boldsymbol{J}_{T} \varphi(\boldsymbol{X} )\varphi( \boldsymbol{X})^{\prime}\boldsymbol{J}_{T} \boldsymbol{y} = 
\boldsymbol{\Lambda}\left(T^{-1} \boldsymbol{F}^{\prime} \boldsymbol{J}_{T} \boldsymbol{F}\right)\left(M^{-1} \boldsymbol{\Phi}^{\prime}  \boldsymbol{\Phi}\right)\left(T^{-1} \boldsymbol{F}^{\prime} \boldsymbol{J}_{T} \boldsymbol{F}\right) \boldsymbol{\beta} \\
 &+\boldsymbol{\Lambda}\left(T^{-1} \boldsymbol{F}^{\prime} \boldsymbol{J}_{T} \boldsymbol{F}\right)\left(M^{-1} \boldsymbol{\Phi}^{\prime}  \boldsymbol{\Phi}\right)\left(T^{-1} \boldsymbol{F}^{\prime} \boldsymbol{J}_{T} \boldsymbol{\eta}\right) 
 +\boldsymbol{\Lambda}\left(T^{-1} \boldsymbol{F}^{\prime} \boldsymbol{J}_{T} \boldsymbol{F}\right)\left(M^{-1} T^{-1} \boldsymbol{\Phi}^{\prime} \boldsymbol{\varepsilon}^{\prime} \boldsymbol{J}_{T} \boldsymbol{F}\right) \boldsymbol{\beta} \\
 &+\boldsymbol{\Lambda}\left(T^{-1} \boldsymbol{F}^{\prime} \boldsymbol{J}_{T} \boldsymbol{F}\right)\left(M^{-1} T^{-1} \boldsymbol{\Phi}^{\prime}  \boldsymbol{\varepsilon}^{\prime} \boldsymbol{J}_{T} \boldsymbol{\eta}\right) 
 +\boldsymbol{\Lambda}\left(M^{-1} T^{-1} \boldsymbol{F}^{\prime} \boldsymbol{J}_{T} \boldsymbol{\varepsilon}  \boldsymbol{\Phi}\right)\left(T^{-1} \boldsymbol{F}^{\prime} \boldsymbol{J}_{T} \boldsymbol{F}\right) \boldsymbol{\beta} \\
& +\boldsymbol{\Lambda}\left(M^{-1} T^{-1} \boldsymbol{F}^{\prime} \boldsymbol{J}_{T} \boldsymbol{\varepsilon}  \boldsymbol{\Phi}\right)\left(T^{-1} \boldsymbol{F}^{\prime} \boldsymbol{J}_{T} \boldsymbol{\eta}\right) 
+\boldsymbol{\Lambda}\left(M^{-1} T^{-2} \boldsymbol{F}^{\prime} \boldsymbol{J}_{T} \boldsymbol{\varepsilon}  \boldsymbol{\varepsilon}^{\prime} \boldsymbol{J}_{T} \boldsymbol{F}\right) \boldsymbol{\beta} \\
 &+ \boldsymbol{\Lambda}\left(M^{-1} T^{-2} \boldsymbol{F}^{\prime} \boldsymbol{J}_{T} \boldsymbol{\varepsilon}  \boldsymbol{\varepsilon}^{\prime} \boldsymbol{J}_{T} \boldsymbol{\eta}\right) 
+\left(T^{-1} \boldsymbol{\omega}^{\prime} \boldsymbol{J}_{T} \boldsymbol{F}\right)\left(M^{-1} \boldsymbol{\Phi}^{\prime}  \boldsymbol{\Phi}\right)\left(T^{-1} \boldsymbol{F}^{\prime} \boldsymbol{J}_{T} \boldsymbol{F}\right) \boldsymbol{\beta} \\
&+\left(T^{-1} \boldsymbol{\omega}^{\prime} \boldsymbol{J}_{T} \boldsymbol{F}\right)\left(M^{-1} \boldsymbol{\Phi}^{\prime}  \boldsymbol{\Phi}\right)\left(T^{-1} \boldsymbol{F}^{\prime} \boldsymbol{J}_{T} \boldsymbol{\eta}\right) 
 +\left(T^{-1} \boldsymbol{\omega}^{\prime} \boldsymbol{J}_{T} \boldsymbol{F}\right)\left(M^{-1} T^{-1} \boldsymbol{\Phi}^{\prime}  \boldsymbol{\varepsilon}^{\prime} \boldsymbol{J}_{T} \boldsymbol{F}\right) \boldsymbol{\beta}\\ 
& +\left(T^{-1} \boldsymbol{\omega}^{\prime} \boldsymbol{J}_{T} \boldsymbol{F}\right)\left(M^{-1} T^{-1} \boldsymbol{\Phi}^{\prime}  \boldsymbol{\varepsilon}^{\prime} \boldsymbol{J}_{T} \boldsymbol{\eta}\right) 
 +\left(M^{-1} T^{-1} \boldsymbol{\omega}^{\prime} \boldsymbol{J}_{T} \boldsymbol{\varepsilon}  \boldsymbol{\Phi}\right)\left(T^{-1} \boldsymbol{F}^{\prime} \boldsymbol{J}_{T} \boldsymbol{F}\right) \boldsymbol{\beta} \\
 &+\left(M^{-1} T^{-1} \boldsymbol{\omega}^{\prime} \boldsymbol{J}_{T} \boldsymbol{\varepsilon}  \boldsymbol{\Phi}\right)\left(T^{-1} \boldsymbol{F}^{\prime} \boldsymbol{J}_{T} \boldsymbol{\eta}\right) 
 +\left(M^{-1} T^{-2} \boldsymbol{\omega}^{\prime} \boldsymbol{J}_{T} \boldsymbol{\varepsilon}  \boldsymbol{\varepsilon}^{\prime} \boldsymbol{J}_{T} \boldsymbol{F}\right) \boldsymbol{\beta}\\ 
&+\left(M^{-1} T^{-2} \boldsymbol{\omega}^{\prime} \boldsymbol{J}_{T} \boldsymbol{\varepsilon}  \boldsymbol{\varepsilon}^{\prime} \boldsymbol{J}_{T} \boldsymbol{\eta}\right) \\
&=  \boldsymbol{\Lambda} \boldsymbol{\Delta}_{F} \mathcal{P} \boldsymbol{\Delta}_{F} \boldsymbol{\beta} + \boldsymbol{O}_{p}\left(\delta_{M T}^{-1}\right)
\end{align*}
 The final line follows directly from Lemma \ref{l1}  and Assumptions \ref{asu2}.\ref{i2.1} and \ref{asu2}.\ref{i2.2}.

\vspace{4mm}
 \noindent  Item 3:
\noindent Let = $\hat{\boldsymbol{F}}_{C,t} = M^{-1} T^{-1} \boldsymbol{Z}^{\prime} \boldsymbol{J}_{T} \varphi(\boldsymbol{X}) \varphi( \boldsymbol{x}_{t}) $. Then, given Lemma \ref{l2}.\ref{l2.1}, standard arguments would imply that  $M^{-2} T^{-3} \boldsymbol{Z}^{\prime} \boldsymbol{J}_{T} \varphi(\boldsymbol{X} )\varphi( \boldsymbol{X})^{\prime} \boldsymbol{J}_{T} \varphi(\boldsymbol{X} )\varphi( \boldsymbol{X})^{\prime} \boldsymbol{J}_{T} \boldsymbol{Z}$  $= \dfrac{\hat{\boldsymbol{F}}_{C} \boldsymbol{J}_{T} {\hat{\boldsymbol{F}}_{C}}^{\prime}}{T}$ \\$=\boldsymbol{\Lambda} \boldsymbol{\Delta}_{F} \mathcal{P} \left( T^{-1}\boldsymbol{F} \boldsymbol{J}_{T} \boldsymbol{F} \right) \mathcal{P} \boldsymbol{\Delta}_{F} \boldsymbol{\Lambda}'  + \boldsymbol{O}_{p}\left(\delta_{M T}^{-1}\right)$. Given Assumption \ref{asu2}.\ref{i2.1}, we have that\\
$\boldsymbol{\Lambda} \boldsymbol{\Delta}_{F} \mathcal{P} \left( T^{-1}\boldsymbol{F} \boldsymbol{J}_{T} \boldsymbol{F} \right) \mathcal{P} \boldsymbol{\Delta}_{F} \boldsymbol{\Lambda}'  $ $=\boldsymbol{\Lambda} \boldsymbol{\Delta}_{F} \mathcal{P} \boldsymbol{\Delta}_{F} \mathcal{P} \boldsymbol{\Delta}_{F} \boldsymbol{\Lambda}^{\prime} + \boldsymbol{O}_{p}\left({T}^{-1/2}\right)$.\\ Therefore, we have that, $M^{-2} T^{-3} \boldsymbol{Z}^{\prime} \boldsymbol{J}_{T} \varphi(\boldsymbol{X} )\varphi( \boldsymbol{X})^{\prime} \boldsymbol{J}_{T} \varphi(\boldsymbol{X} )\varphi( \boldsymbol{X})^{\prime} \boldsymbol{J}_{T} \boldsymbol{Z}$ $=\boldsymbol{\Lambda} \boldsymbol{\Delta}_{F} \mathcal{P} \boldsymbol{\Delta}_{F} \mathcal{P} \boldsymbol{\Delta}_{F} \boldsymbol{\Lambda}^{\prime} + \boldsymbol{O}_{p}\left(\delta_{M T}^{-1}\right) + \boldsymbol{O}_{p}\left({T}^{-1/2}\right) =$  $\boldsymbol{O}_{p}\left(\delta_{M T}^{-1}\right)$.

\vspace{5mm}
\noindent \textbf{Theorem 1} If Assumption \ref{asu1}-\ref{asu5} hold, we have 
    \[
\hat{\boldsymbol{F}}_{t}- \boldsymbol{H}_f \boldsymbol{f}_{t} = \boldsymbol{O}_p(\delta_{M T}^{-1} )
\]\\
where $\boldsymbol{H}_{f} \equiv \hat{\boldsymbol{F}}_{A} \hat{\boldsymbol{F}}_{B}^{-1}\boldsymbol{\Lambda}_f \boldsymbol{\Delta}_{f} $ \\
$\hat{\boldsymbol{F}}_{A} = T^{-1} \boldsymbol{Z}^{\prime} \boldsymbol{J}_{T} \boldsymbol{Z}$ and\\ $\hat{\boldsymbol{F}}_{B} =M^{-1} T^{-2} \boldsymbol{Z}^{\prime} \boldsymbol{J}_{T} \varphi(\boldsymbol{X}) \varphi(\boldsymbol{X}^{\prime}) \boldsymbol{J}_{T} \boldsymbol{Z}$

\noindent \textit{Proof:} 
\begin{align*}
    \hat{\boldsymbol{F}}_{t} & =T^{-1} \boldsymbol{Z}^{\prime} \boldsymbol{J}_{T} \boldsymbol{Z}\left(M^{-1} T^{-2} \boldsymbol{Z}^{\prime} \boldsymbol{J}_{T} \varphi(\boldsymbol{X}) \varphi(\boldsymbol{X})^{\prime} \boldsymbol{J}_{T} \boldsymbol{Z}\right)^{-1} M^{-1} T^{-1} \boldsymbol{Z}^{\prime} \boldsymbol{J}_{T} \varphi(\boldsymbol{X}) \varphi( \boldsymbol{x}_{t})\\
    &= \hat{\boldsymbol{F}}_{A} \hat{\boldsymbol{F}}_{B}^{-1}\left( \boldsymbol{\Lambda} \boldsymbol{\Delta}_{F} \mathcal{P} \boldsymbol{F}_{t} + \boldsymbol{O}_{p}\left(\delta_{M T}^{-1}\right)\right)\\
    &= \hat{\boldsymbol{F}}_{A} \hat{\boldsymbol{F}}_{B}^{-1}\boldsymbol{\Lambda} \boldsymbol{\Delta}_{F} \mathcal{P} \boldsymbol{F}_{t} + \boldsymbol{O}_{p}\left(\delta_{M T}^{-1}\right)\\
     &= \hat{\boldsymbol{F}}_{A} \hat{\boldsymbol{F}}_{B}^{-1}\boldsymbol{\Lambda}_f \boldsymbol{\Delta}_{f} \boldsymbol{f}_{t} + \boldsymbol{O}_{p}\left(\delta_{M T}^{-1}\right)\\
    &= \boldsymbol{H}_f \boldsymbol{f}_{t} + \boldsymbol{O}_p(\delta_{M T}^{-1})
\end{align*}

\noindent The second equality follows from Lemma \ref{l2}.\ref{l2.1}. The second last equality follows from the same reasoning as in the Proof of Theorem 1 in \cite{kelly}, noting that $\mathcal{P}_1$ as described in their setup is an identity matrix by Assumption \ref{asu4}. Final equality uses the definition of $\boldsymbol{H}_{f}$.

\noindent \textbf{Theorem 2} If Assumption \ref{asu1}-\ref{asu5} hold, we have 
   $$\hat{\boldsymbol{\beta}}-\boldsymbol{G}_{\beta}\boldsymbol{\beta}_f =  \boldsymbol{O}_p(\delta_{M T}^{-1} ).$$
   where $\boldsymbol{G}_{\beta} \equiv \hat{\boldsymbol{\beta}}_{1} ^{-1} {\hat{\boldsymbol{\beta}}_{2}} \left[\boldsymbol{\Lambda}_f \boldsymbol{\Delta}_{f}^3  \boldsymbol{\Lambda}_f^{\prime}\right]^{-1} \boldsymbol{\Lambda}_f \boldsymbol{\Delta}_{f}^2 $, where\\
$\hat{\boldsymbol{\beta}}_{1}=\hat{\boldsymbol{F}}_{A}$ and $\hat{\boldsymbol{\beta}}_{2}=\hat{\boldsymbol{F}}_{B}$, and $\boldsymbol{\Delta}_{f}$ is defined in the paragraph preceding Theorem \ref{t1}. ,\\

\vspace{3mm}
\noindent \textit{Proof:} 
\begin{align*}
    \hat{\boldsymbol{\beta}} & = \left( T^{-1} \boldsymbol{Z}^{\prime} \boldsymbol{J}_{T} \boldsymbol{Z}\right)^{-1} M^{-1} T^{-2} \boldsymbol{Z}^{\prime} \boldsymbol{J}_{T} \varphi(\boldsymbol{X})  \varphi(\boldsymbol{X})^{\prime} \boldsymbol{J}_{T} \boldsymbol{Z} \\ 
    & \times \left(M^{-2} T^{-3}\boldsymbol{Z}^{\prime} \boldsymbol{J}_{T} \varphi(\boldsymbol{X}) \varphi(\boldsymbol{X})^{\prime} \boldsymbol{J}_{T} \varphi(\boldsymbol{X})  \varphi(\boldsymbol{X})^{\prime} \boldsymbol{J}_{T} \boldsymbol{Z}\right)^{-1} M^{-1} T^{-2} \boldsymbol{Z}^{\prime} \boldsymbol{J}_{T} \varphi(\boldsymbol{X})  \varphi(\boldsymbol{X})^{\prime} \boldsymbol{J}_{T} \boldsymbol{y}\\
    & =  \hat{\boldsymbol{\beta}}_{1} ^{-1} {\hat{\boldsymbol{\beta}}_{2}} \left( \boldsymbol{\Lambda} \boldsymbol{\Delta}_{F} \mathcal{P} \boldsymbol{\Delta}_{F} \mathcal{P} \boldsymbol{\Delta}_{F} \boldsymbol{\Lambda}^{\prime} + \boldsymbol{O}_{p}\left(\delta_{M T}^{-1}\right) \right)^{-1} 
    \left(\boldsymbol{\Lambda} \boldsymbol{\Delta}_{F} \mathcal{P} \boldsymbol{\Delta}_{F} \boldsymbol{\beta} + \boldsymbol{O}_{p}\left(\delta_{M T}^{-1}\right) \right)\\
    & =  \hat{\boldsymbol{\beta}}_{1} ^{-1} {\hat{\boldsymbol{\beta}}_{2}} \left[ \left( \boldsymbol{\Lambda} \boldsymbol{\Delta}_{F} \mathcal{P} \boldsymbol{\Delta}_{F} \mathcal{P} \boldsymbol{\Delta}_{F} \boldsymbol{\Lambda}^{\prime} \right)^{-1} -  \left( \boldsymbol{\Lambda} \boldsymbol{\Delta}_{F} \mathcal{P} \boldsymbol{\Delta}_{F} \mathcal{P} \boldsymbol{\Delta}_{F} \boldsymbol{\Lambda}^{\prime} \right)^{-1}\boldsymbol{O}_{p}\left(\delta_{M T}^{-1}\right) \left( \boldsymbol{O}_{p}(1) + \boldsymbol{O}_{p}\left(\delta_{M T}^{-1}\right)\right)^{-1} \right] \times \\
    &\left(\boldsymbol{\Lambda} \boldsymbol{\Delta}_{F} \mathcal{P} \boldsymbol{\Delta}_{F} \boldsymbol{\beta} + \boldsymbol{O}_{p}\left(\delta_{M T}^{-1}\right) \right)\\
    & =  \hat{\boldsymbol{\beta}}_{1} ^{-1} {\hat{\boldsymbol{\beta}}_{2}} \left[\boldsymbol{\Lambda} \boldsymbol{\Delta}_{F} \mathcal{P} \boldsymbol{\Delta}_{F} \mathcal{P} \boldsymbol{\Delta}_{F} \boldsymbol{\Lambda}^{\prime}\right]^{-1} \boldsymbol{\Lambda} \boldsymbol{\Delta}_{F} \mathcal{P} \boldsymbol{\Delta}_{F}{\boldsymbol{\beta}} + \boldsymbol{O}_{p}\left(\delta_{M T}^{-1}\right)\\
     & =  \hat{\boldsymbol{\beta}}_{1} ^{-1} {\hat{\boldsymbol{\beta}}_{2}} \left[\boldsymbol{\Lambda}_f \boldsymbol{\Delta}_{f}^3  \boldsymbol{\Lambda}_f^{\prime}\right]^{-1} \boldsymbol{\Lambda}_f \boldsymbol{\Delta}_{F}^2 {\boldsymbol{\beta}}_f + \boldsymbol{O}_{p}\left(\delta_{M T}^{-1}\right)\\
    & = \boldsymbol{G}_{\beta}\boldsymbol{\beta} +  \boldsymbol{O}_p(\delta_{M T}^{-1} )
\end{align*}
where the second equality employs Lemma \ref{l2}.\ref{l2.2} and \ref{l2}.\ref{l2.3}. The third equality uses the fact that for any invertible matrices $\boldsymbol{A} $ and $\boldsymbol{A} + \boldsymbol{B}$ we have  $\left( \boldsymbol{A} + \boldsymbol{B} \right)^{-1} =
\boldsymbol{A}^{-1} - \boldsymbol{A}^{-1}\boldsymbol{B}\left( \boldsymbol{A} + \boldsymbol{B} \right)^{-1} $, which in our case implies that,\\
$\left(\boldsymbol{\Lambda} \boldsymbol{\Delta}_{F} \mathcal{P} \boldsymbol{\Delta}_{F} \mathcal{P} \boldsymbol{\Delta}_{F} \boldsymbol{\Lambda}^{\prime} + \boldsymbol{O}_p(\delta_{M T}^{-1} )\right)^{-1} =$\\
$\left(\boldsymbol{\Lambda} \boldsymbol{\Delta}_{F} \mathcal{P} \boldsymbol{\Delta}_{F} \mathcal{P} \boldsymbol{\Delta}_{F} \boldsymbol{\Lambda}^{\prime} \right)^{-1} -\left(\boldsymbol{\Lambda} \boldsymbol{\Delta}_{F} \mathcal{P} \boldsymbol{\Delta}_{F} \mathcal{P} \boldsymbol{\Delta}_{F} \boldsymbol{\Lambda}^{\prime} \right)^{-1}\boldsymbol{O}_p(\delta_{M T}^{-1})  \left(\boldsymbol{\Lambda} \boldsymbol{\Delta}_{F} \mathcal{P} \boldsymbol{\Delta}_{F} \mathcal{P} \boldsymbol{\Delta}_{F} \boldsymbol{\Lambda}^{\prime} + \boldsymbol{O}_p(\delta_{M T}^{-1})\right)^{-1} . $
The last equality uses the definition of $\boldsymbol{G}_{\beta}$. The second-to-last equality follows the logic in the proof of Theorem 1 of \cite{kelly}. Under Assumption \ref{asu5} the normalization assumption \ref{asu4}, we can partition
\[
\boldsymbol{\Lambda} \boldsymbol{\Delta}_{F} \mathcal{P} \boldsymbol{\Delta}_{F} = \left[ \boldsymbol{\Lambda}_f \boldsymbol{\Delta}_f \mathcal{P}_f \boldsymbol{\Delta}_f, \ \mathbf{0} \right],
\]
where $\mathcal{P}_f$ is the covariance matrix of the relevant factors and equals the identity matrix under Assumption \ref{asu4}. Consequently,
\[
\boldsymbol{\Lambda} \boldsymbol{\Delta}_F \mathcal{P} \boldsymbol{\Delta}_F \mathcal{P} \boldsymbol{\Delta}_F \boldsymbol{\Lambda}' = \boldsymbol{\Lambda}_f \boldsymbol{\Delta}_f \mathcal{P}_f \boldsymbol{\Delta}_f \mathcal{P}_f \boldsymbol{\Delta}_f \boldsymbol{\Lambda}_f' = \boldsymbol{\Lambda}_f \boldsymbol{\Delta}_f^3 \boldsymbol{\Lambda}_f',
\]
and similarly,
\[
\boldsymbol{\Lambda} \boldsymbol{\Delta}_F \mathcal{P} \boldsymbol{\Delta}_F \boldsymbol{\beta} = \boldsymbol{\Lambda}_f \boldsymbol{\Delta}_f \mathcal{P}_f \boldsymbol{\Delta}_f \boldsymbol{\beta}_f = \boldsymbol{\Lambda}_f \boldsymbol{\Delta}_f^2 \boldsymbol{\beta}_f.
\]

\noindent \textbf{Theorem 3}  If Assumption \ref{asu1}-\ref{asu5} hold, we have 
 $$ \hat{y}_{t+h}- \mathbb{E}_{t} y_{t+h} = O_p(\delta_{M T}^{-1}) $$

 \vspace{3mm}
 \noindent \textit{Proof:} 
 \begin{align*}
     \hat{y}_{t+h} &= \bar{y} + \boldsymbol{J}_{T}\hat{\boldsymbol{F}}_{T}^{\prime} \hat{\boldsymbol{\beta}}\\
     & = \beta_{0}+\Bar{\boldsymbol{f}}^{\prime} \boldsymbol{\beta}_f+ O_p(T^{-1/2}) + 
     \left( \boldsymbol{H}_f \boldsymbol{f}_{t} + \boldsymbol{O}_p(\delta_{M T}^{-1}) \right)^{\prime}
     \left( \boldsymbol{G}_{\beta}\boldsymbol{\beta}_f +  \boldsymbol{O}_p(\delta_{M T}^{-1} )\right)\\
     & = \beta_{0}+\Bar{\boldsymbol{f}}^{\prime} \boldsymbol{\beta}_f +\left( \boldsymbol{f}_{t} - \Bar{\boldsymbol{f}} \right)^{\prime}\boldsymbol{H}_f^{\prime} \boldsymbol{G}_{\beta}\boldsymbol{\beta}_f +  \boldsymbol{O}_p(\delta_{M T}^{-1} )\\
     & = \beta_{0}+\Bar{\boldsymbol{f}}^{\prime} \boldsymbol{\beta}_f +\left( \boldsymbol{f}_{t} - \Bar{\boldsymbol{f}} \right)^{\prime}\boldsymbol{\beta}_f +  \boldsymbol{O}_p(\delta_{M T}^{-1} )\\
     & = \beta_{0} + \boldsymbol{f}_{t}^{\prime}\boldsymbol{\beta}_f + \boldsymbol{O}_p(\delta_{M T}^{-1})\\
     & = \mathbb{E}_{t} y_{t+h} + O_p(\delta_{M T}^{-1}) 
 \end{align*}
The second equality follows from lemma \ref{l1}.\ref{2.15}. The fourth equality follows if  $\boldsymbol{H}_f^{\prime} \boldsymbol{G}_{\beta}$ is an identity matrix. 

\subsection{Mercer's Theorem}\label{apx_mercer_theorem}
Suppose $\mathcal{X} \subseteq \mathbb{R}^d$ is compact and kernel function  $\mathcal{K}: \mathcal{X} \times \mathcal{X} \rightarrow \mathbb{R}$ is continuous, satisfying the following conditions, 
$$
\int_{\mathrm{y}} \int_{\mathrm{x}} \mathcal{K}^2(\mathrm{x}, \mathrm{y}) d \mathrm{x} d \mathrm{y}<\infty \quad \text { and } \int_{\mathrm{y}} \int_{\mathrm{x}} h(\mathrm{x}) \mathcal{K}(\mathrm{x}, \mathrm{y}) h(\mathrm{y}) d \mathrm{x} d \mathrm{y} \geq 0, \quad \forall h \in L^2(\mathcal{X}),
$$
where $L^2(\mathcal{X})=\left\{h: \int h^2(\mathrm{s}) d \mathrm{s}<\infty\right\}$, then there exist functions $\left\{\varphi_i(\cdot) \in L^2(\mathcal{X}), i=1,2, \ldots\right\}$   and  non-negative coefficients $\theta_1 \geq \theta_2 \geq \ldots \geq 0$ which together forms an orthonormal system in $L^2(\mathcal{X})$, i.e. $\left\langle\varphi_i, \varphi_j\right\rangle_{L^2(\mathcal{X})}=\int \varphi_i(\mathrm{x}) \varphi_j(\mathrm{x}) d \mathrm{x}=\mathbb{I}_{\{i=j\}}$, such that
$$
\mathcal{K}(\mathrm{x}, \mathrm{y})=\sum_{i=1}^{\infty} \theta_i \varphi_i(\mathrm{x}) \varphi_i(\mathrm{y}), \quad \forall \mathrm{x}, \mathrm{y} \in \mathcal{X}
$$

\end{document}